\documentclass[twocolumn,aps,prd,floatfix,preprintnumbers,a4paper,nofootinbib,superscriptaddress,10pt]{revtex4-1} 
\usepackage{lipsum}
\usepackage{multirow}
\usepackage{import}
\usepackage{tabularx}
\usepackage[titletoc,title]{appendix}

\usepackage{graphicx}

\usepackage{amsmath}
\usepackage{amssymb}
\usepackage{amsfonts}
\usepackage{mathtools}
\usepackage{bm}
\usepackage{import}
\usepackage{placeins}

\usepackage[utf8]{inputenc}
\usepackage{newtxtext}

\usepackage[table]{xcolor}
\usepackage{url}
\usepackage[colorlinks, pdfborder={0 0 0}]{hyperref}
\definecolor{LinkColor}{rgb}{0.75, 0, 0}
\definecolor{CiteColor}{rgb}{0.75, 0, 0}
\definecolor{UrlColor}{rgb}{0, 0, 0.75}
\hypersetup{linkcolor=LinkColor}
\hypersetup{citecolor=CiteColor}
\hypersetup{urlcolor=UrlColor}

\usepackage{rotating}
\usepackage{booktabs}

\usepackage{dcolumn}
\newcolumntype{d}{D{.}{.}{-1}}
\newcolumntype{b}[1]{D{(}{(}{#1}}

\usepackage{xspace}
\usepackage[normalem]{ulem}

\newcommand{\UIB}{Departament de Física, Universitat de les Illes Balears, IAC3 – IEEC, Crta. Valldemossa km 7.5, E-07122 Palma, Spain}
\newcommand{\Soton}{Mathematical Sciences \& STAG Research Centre, University of Southampton, Southampton, SO17 1BJ, United Kingdom}
\newcommand{\Portsmouth}{Institute of Cosmology \& Gravitation, University of Portsmouth, Portsmouth, United Kingdom}
\newcommand{\Cardiff}{Gravity Exploration Institute, School of Physics and Astronomy, Cardiff University, Cardiff CF24 3AA, United Kingdom}
\newcommand{\Zurich}{Physik-Institut, Universit\"at Z\"urich, Winterthurerstrasse 190, 8057 Z\"urich, Switzerland}
\newcommand{\KCL}{King's  College  London,  Strand,  London  WC2R  2LS,  United Kingdom}
\newcommand{\AEI}{Max Planck Institute for Gravitational Physics (Albert Einstein Institute),
Callinstrasse 38, D-30167 Hannover, Germany}
\newcommand{\Leibniz}{Leibniz University Hannover, 30167 Hannover, Germany}
\newcommand{\ICE}{Institut de Ci\`encies de l’Espai (ICE, CSIC), Campus UAB, Carrer de Can Magrans s/n, 08193 Cerdanyola del Vall\`es, Spain}
\newcommand{\Caltech}{Theoretical Astrophysics Group, California Institute of Technology, Pasadena, CA 91125, U.S.A.}

\definecolor{darkgreen}{RGB}{0,153,0}
\definecolor{epurple}{RGB}{216,131,183}
\definecolor{gold}{HTML}{edbb00}
\newcommand{\st}{\textsc{SpinTaylor}\xspace}

\usepackage[acronym,shortcuts]{glossaries}

\newacronym{bbh}{BBH}{binary black hole}

\begin{document}

\def\xas#1{\texttt{PhenomXAS}}
\def\xhm#1{\texttt{PhenomXHM}}
\def\xphm#1{\texttt{PhenomXPHM}}
\def\xcp#1{\texttt{PhenomXHM-CP}}
\def\gw#1{gravitational wave#1}
\def\nr#1{numerical relativity
 (NR)#1\gdef\nr{NR}}
\def\bh#1{black-hole
 (BH)#1\gdef\bh{BH}}
 \def\bbh#1{binary black hole#1
  (BBH#1)\gdef\bbh{BBH}}
 \def\qnm#1{Quasinormal Modes
    (QNM)#1\gdef\qnm{QNM}}
\def\oed#1{optimal emission direction#1}
\def\pn#1{post-Newtonian (PN)#1\gdef\pn{PN}}
\def\imr#1{inspiral-merger-ringdown (IMR)#1\gdef\imr{IMR}}
   \def\eob#1{effective-one-body
      (EOB)#1\gdef\eob{EOB}}
\def\td#1{time-domain (TD)#1\gdef\td{TD}}
\def\fd#1{frequency-domain (FD)#1\gdef\fd{FD}}

\title{PhenomXPNR: An improved gravitational wave model linking precessing inspirals and NR-calibrated merger-ringdown}


\author{Eleanor Hamilton}
\affiliation{\UIB}
\author{Marta Colleoni}
\affiliation{\UIB}

\author{Jonathan E. Thompson}
\affiliation{\Soton}
\affiliation{\Caltech}
\author{Charlie Hoy}
\affiliation{\Portsmouth}

\author{Anna Heffernan}
\affiliation{\UIB}
\author{Meryl Kinnear}
\affiliation{\Cardiff}
\author{Jorge Valencia}
\affiliation{\UIB}
\author{Felip A. Ramis Vidal}
\affiliation{\UIB}

\author{Cecilio Garc\'ia-Quir\'os}
\affiliation{\Zurich}
\author{Shrobana Ghosh}
\affiliation{\AEI}
\affiliation{\Leibniz}
\author{Lionel London}
\affiliation{\KCL}

\author{Mark Hannam}
\affiliation{\Cardiff}
\author{Sascha Husa}
\affiliation{\ICE}
\affiliation{\UIB}

\begin{abstract}
We present the frequency-domain quasi-circular precessing binary-black-hole model \textsc{PhenomXPNR}.
This model combines the most precise available post-Newtonian description of the evolution of the precession dynamics through inspiral  with merger-ringdown model informed by numerical relativity.
This, along with a phenomenological model of the dominant multipole asymmetries, results in the most accurate and complete representation of the physics of precessing binaries natively in the frequency-domain to date.
All state-of-the-art precessing models show bias when inferring binary parameters in certain regions of the parameter space.
We demonstrate that the developments presented ensure that for some precessing systems \textsc{PhenomXPNR} shows the least degree of bias.
Further, as a phenomenological, frequency-domain model, \textsc{PhenomXPNR} remains one of the most computationally efficient models available and is therefore well-suited to the era of gravitational-wave astronomy with its ever growing rate of detected signals.
\end{abstract}

\date{\today}

\maketitle

\section{Introduction}
\label{sec:intro}

Since 2015 $\sim$100 \gw{} signals have been detected by the LIGO-Virgo-KAGRA (LVK) network~\cite{LIGOScientific:2021vkt}, most of them from merging black holes. 
The black holes' properties are measured by comparing the detector data to theoretical signal models. 
The measurement accuracy is determined by the statistical error due to detector noise, but also the accuracy of 
the models themselves. As detector sensitivity increases, so does the required model accuracy. 
Detector upgrades have led to roughly a factor of two increase in sensitivity over the last decade~\cite{abbott2020prospects}, with a further factor of two for current generation ground-based detectors projected into the 2030s, and another order of magnitude improvement expected in 
next-generation ground-based detectors by 2040~\cite{Punturo:2010zza, Hild:2010id, LIGOScientific:2016wof, Reitze:2019iox}. 
In parallel, successive improvements in 
waveform models have seen improved tuning to analytic approximations and \nr{} simulations, 
and inclusion of more detailed physical effects. 

In this paper, we focus on the modelling of spin precession effects~\cite{Apostolatos:1994mx, Kidder:1995zr}. 
This is challenging due to the complex phenomenology and high dimensionality of such systems. There is a high computational cost required to densely cover the seven-dimensional parameter space (mass ratio and two spin vectors) with long \nr{} simulations in order to understand their detailed characteristics. 
Waveform models that describe precessing spins have however already been used to analyse \gw{} signals since the first \gw{} detection~\cite{LIGOScientific:2016aoc,LIGOScientific:2016vlm}. 
This has been made possible by the twisting-up approximation~\cite{Schmidt:2010it,Schmidt:2012rh, OShaughnessy:2011pmr, Boyle:2011gg}, which provides an approximate map between precessing and non-precessing waveforms.
Many precessing \gw{} models~\cite{Pratten:2020fqn, Colleoni:2024knd, Estelles:2021gvs, Ramos-Buades:2023ehm, Gamba:2024cvy} utilise information from \pn{} and \bh{} perturbation theory to describe this mapping for the entire \imr{} signal without employing information from \nr{} in order to ensure an accurate merger-ringdown prescription.

Increasing detector sensitivity results in an increased observable volume of the universe and a consequent increase in the number of detections.
Analysing the large number of detected signals anticipated in this era of gravitational wave astronomy requires both highly accurate and highly computationally efficient models.
To date, state-of-the-art models have used three main modelling approaches: phenomenological waveform (Phenom) models~\cite{Pratten:2020fqn, Garcia-Quiros:2020qpx, Pratten:2020ceb,  Estelles:2021gvs, Colleoni:2024knd, Thompson:2023ase}, the \eob{} paradigm ~\cite{Khalil:2023kep, Pompili:2023tna, Ramos-Buades:2023ehm, vandeMeent:2023ols, Mihaylov:2023bkc, Gamba:2024cvy} and surrogate models~\cite{Varma:2018mmi, Varma:2019csw, Yoo:2022erv}.
The Phenom and EOBNR models use semi-analytic approximations to describe the \gw{} signal during the black holes' inspiral and \nr{} simulations to inform a model of the merger and ringdown.
The surrogate models are trained purely on \nr{}  or hybrid waveforms to accurately model the signal, without analytic approximations over the binary parameter space of the training data.

The above models can be further divided into two broad categories; those where the modelling takes place natively in the time domain, and those where it occurs in the frequency domain.
Fast \fd{} models constructed in the Phenom approach, such as the model presented here, are typically more convenient for analysis of detected \gw{} signals from compact binary sources. The analysis of a single \gw{} signal requires millions of likelihood evaluations: under the assumption of stationary and Gaussian noise, the noise covariance matrix of GW detectors becomes diagonal in the Fourier domain and the likelihood can be recast into a computationally efficient form known as the Whittle likelihood~\cite{Whittle}. This assumption, combined with the circulant hypothesis, enables the use of classical sampling methods to perform Bayesian parameter estimation in the \fd{}, and facilitates the marginalisation of the likelihood over nuisance parameters~\cite{Thrane_2019}. 

The PhenomX family of models~\cite{Pratten:2020fqn, Garcia-Quiros:2020qpx, Pratten:2020ceb, Colleoni:2024knd, Thompson:2023ase} constitutes the current generation of \fd{} models for the \gw{} signal from coalescing binaries.
These have been used e.g. in the GWTC-3 catalog of LVK observations~\cite{KAGRA:2021vkt}.
The underlying aligned-spin models (\textsc{PhenomXAS}~\cite{Pratten:2020fqn} for the dominant quadrupole modes, which is extended to other spherical harmonics in \textsc{PhenomXHM}~\cite{Garcia-Quiros:2020qpx}) are calibrated to \nr{} simulations of non-precessing (aligned-spin) binaries up to mass ratios of 1:18, and black-hole spins up to 0.995 in the equal mass limit and up to 0.85 at higher mass ratios, with input from \pn{} and \eob{} results for the inspiral, and perturbation-theory results to inform the ringdown and the extreme-mass-ratio limit. 
In the \textsc{PhenomXO4a}~\cite{Thompson:2023ase} extension of these models, spin-precession effects are captured during the inspiral by a multi-scale analysis (MSA) model~\cite{Chatziioannou:2016ezg, Chatziioannou:2017tdw}, and through the merger and ringdown are tuned to single-spin NR simulations up to 
mass ratios of 1:8 and spins up to 0.8~\cite{Hamilton:2021pkf}. 
\textsc{PhenomXO4a} also includes a model of the dominant contribution to the 
multipole asymmetry that leads to out-of-plane recoil of the remnant black hole~\cite{Ghosh:2023mhc}. 
In the \textsc{PhenomXPHM-SpinTaylor}~\cite{Colleoni:2024knd} model
there is no \nr{} tuning of merger-ringdown precession effects, but the inspiral spin precession is described by evolving the 
\st equations, which include higher-order spin corrections than the MSA treatment, and are more robust across
the \bbh{} parameter space.

In this paper we present a new Phenom model, \textsc{PhenomXPNR}, which combines the complementary features of \textsc{PhenomXPHM-SpinTaylor} and \textsc{PhenomXO4a}; the seamless integration of the two models over the full coalescence is achieved through a mapping between the double-spin inspiral precession dynamics and the single-spin NR-tuned model through the merger and ringdown, which we will describe below. 
This prescription is sufficient to capture the broad features of precession effects through merger, with the caveat that one must first identify the appropriate spin values at merger; the latter are taken from the numerical evolution of the PN spin-precession equations previously implemented in \textsc{PhenomXPHM-SpinTaylor}.

We find that \textsc{PhenomXPNR} is the most accurate \imr{} \fd{} model for non-eccentric \bbh{} \gw{} signals, and
as such it supersedes all of its predecessors.
Additionally, it achieves comparable accuracy to complete \imr{} \td{} models for precessing signals, with particular improvement for heavy mass, approximately face-on binaries.

\textsc{PhenomXPNR} is the first semi-analytic model to incorporate NR calibration in the precessing sector and the effect of mode-asymmetry while retaining the computational efficiency required by large scale parameter estimation studies, as e.g.~\gw{} event catalogs. This is aided by the choice to work in the \fd{}, where it is straightforward to include information from all the spherical harmonic modes above a fixed cutoff frequency in the likelihood calculation. In contrast, \td{} waveform models require careful adjustments of their starting frequency to ensure that all non-negligible harmonics are in the detector's frequency band. Finally, the model is available as open source code within LALSuite, and has been reviewed by the LVK.

After establishing the notation and conventions in Sec.~\ref{sec:conventions}, we summarise the main technical aspects of the model in Sec.~\ref{sec:model}. Model accuracy, benchmarks and applications to source property measurements are presented in Sec.~\ref{sec:performance}, followed by our conclusions in Sec.~\ref{sec:conclusions}.

\section{Notation and conventions}
\label{sec:conventions}

In the absence of orbital eccentricity, black hole binaries with misaligned spins are described by eight
intrinsic parameters: the component masses $m_i$ and spin vectors $ \mathbf{S}_i$.
The total mass of the binary is given by $M = m_1 + m_2$ and serves as a scale parameter.
We define the mass ratios $q = m_2/m_1 \leq 1$ and $Q = 1/q \geq 1$ in addition to the symmetric mass ration $\eta = m_1 m_2 / \left(m_1+m_2\right)^2$.
The dimensionless spins are given by $\boldsymbol{\chi}_i = \mathbf{S}_i/m_i^2$, where $|\boldsymbol{\chi}_i|\in\left[0,1\right]$ respects the Kerr limit.

It is often useful to decompose the spins into components parallel and perpendicular to the Newtonian orbital angular momentum $\mathbf{L}$; $\chi^\parallel_i = \boldsymbol{\chi}_i\cdot\hat{\mathbf{L}}$ and $\chi^\perp_i = \boldsymbol{\chi}_i - \chi^\parallel_i\hat{\mathbf{L}}$, where the hat indicates the unit vector. 
It can also be useful to consider the combinations of the spin components as these dominate the effect of the spins on the waveform.
The dominant contribution to the aligned-spin (the spins parallel to the orbital angular momentum) behaviour is parameterised by~\cite{Ajith:2011}
\begin{align} 
   \chi_\mathrm{eff} = {}& \frac{m_1 \chi_1^\parallel + m_2 \chi_2^\parallel}{m_1+m_2},
\end{align}
while the effective precession spin~\cite{Schmidt:2015}
\begin{align} 
   \chi_\mathrm{p} = {}& S_\mathrm{p}/m_1^2,
\end{align}
where $S_\mathrm{p} = \frac{1}{A_1}\max\left(A_1 S_1^\perp, A_2 S_2^\perp\right), A_1 = 2+3m_2/(2m_1)$ and $A_2 = 2+3m_1/(2m_2)$, captures the dominant contribution to the binary spin in the orbital plane. 

\section{Model outline}
\label{sec:model}

\textsc{IMRPhenomXPNR} is a complete \imr{} gravitational wave model for quasi-circular precessing binaries of arbitrary total mass.
The model uses results from \pn{} theory during the inspiral, while the merger-ringdown portion of the signal is calibrated to \nr{}.
Precession effects complicate the modelling of \gw{} signals, introducing oscillations into the amplitude and phase of the waveform. 
We therefore simplify modelling efforts by decomposing the waveform into the model for an aligned-spin binary and a frequency-dependent rotation which tracks the precession of the binary, as first proposed in Ref.~\cite{Schmidt:2010it}.
This rotation is described by the Euler angles $\{\alpha,\beta,\gamma\}$~\cite{Schmidt:2012rh, OShaughnessy:2011pmr, Boyle:2011gg}, which transform from the co-precessing frame to an inertial frame in which the total angular momentum of the binary is along the $z$-direction.
The co-precessing model is based on the aligned-spin model \textsc{PhenomXHM}~\cite{Garcia-Quiros:2020qpx}, with modifications to the merger and ringdown.
Additionally, multipole asymmetries are included in \textsc{PhenomXPNR}.
This co-precessing model follows the same procedure as presented in \textsc{PhenomXO4a}~\cite{Thompson:2023ase}.
The model for the angles $\alpha$ and $\beta$ employs the \st~\cite{Colleoni:2024knd} prescription during the inspiral and a phenomenological description that has been calibrated to \nr{} during merger and ringdown~\cite{Hamilton:2021pkf}.

The \pn{} inspiral prescription accounts for two-spin effects, while the calibration of the precessing sector to \nr{} is performed against a catalogue of 80 single-spin precessing simulations where the spin is placed on the larger black hole~\cite{Hamilton:2023qkv}. We therefore require a mapping between the complete two-spin configuration and the single-spin calibration parameters.We outline the model in the following sections, recapping previous frameworks and highlighting recent developments.

\subsection{Single spin mapping}
\label{sec: single spin map}

We employ a single-spin mapping to go between the full two-spin description of a binary to the single-spin characterisation of the merger-ringdown used in the calibration of the model against \nr{}.
In the single-spin case, we characterise the binary's spins by the dimensionless magnitude $\chi$ and spin inclination $\theta_\mathrm{LS}$ of the larger black hole, taking the smaller black hole to be non-spinning and ignoring in-plane spin rotation.
We calculate these mapped merger-ringdown parameters at the frequency at which the merger-ringdown angles are computed. This transition frequency does not have a closed-form expression as a function of the binary's parameters, since it is related to the stopping frequency of the PN spin-precession equations. It is at this frequency that we compute the exact equivalent single-spin configuration required by the calibrated model for the Euler angles. 
This is given by 
\begin{align} 
   \chi = {}& \sqrt{\chi_\parallel^2 +\chi_\perp^2}, \\
   \cos\theta_\mathrm{LS} = {}& \frac{\chi_\parallel}{\chi},
\end{align}
where 
\begin{align} 
   \chi_\parallel = {}& \frac{M\chi_\mathrm{eff}}{m_1}, \\
   \chi_\perp = {}& \frac{\left| \mathbf{S}^\perp_1 + \mathbf{S}^\perp_2 \right|}{m_1^2}, \label{eqn: chi perp}
\end{align}
and $\mathbf{S}^\perp_i$ are the in-plane spin components. 
We employ $\chi_\mathrm{eff}$ to capture the dominant behaviour of the aligned-spin contribution, which changes monotonically with time, but use the sum of the spin angular momenta at the transition frequency since we wish to capture the instantaneous value of the in-plane spin component, which varies sinusoidally as the spins combine constructively and destructively.
Note that for close-to-equal mass binaries, $\chi$ can exceed the Kerr limit, since it is just a mapped parameter rather than a physical quantity. 

\begin{figure}[t] 
   \centering
   \includegraphics[width=0.48\textwidth]{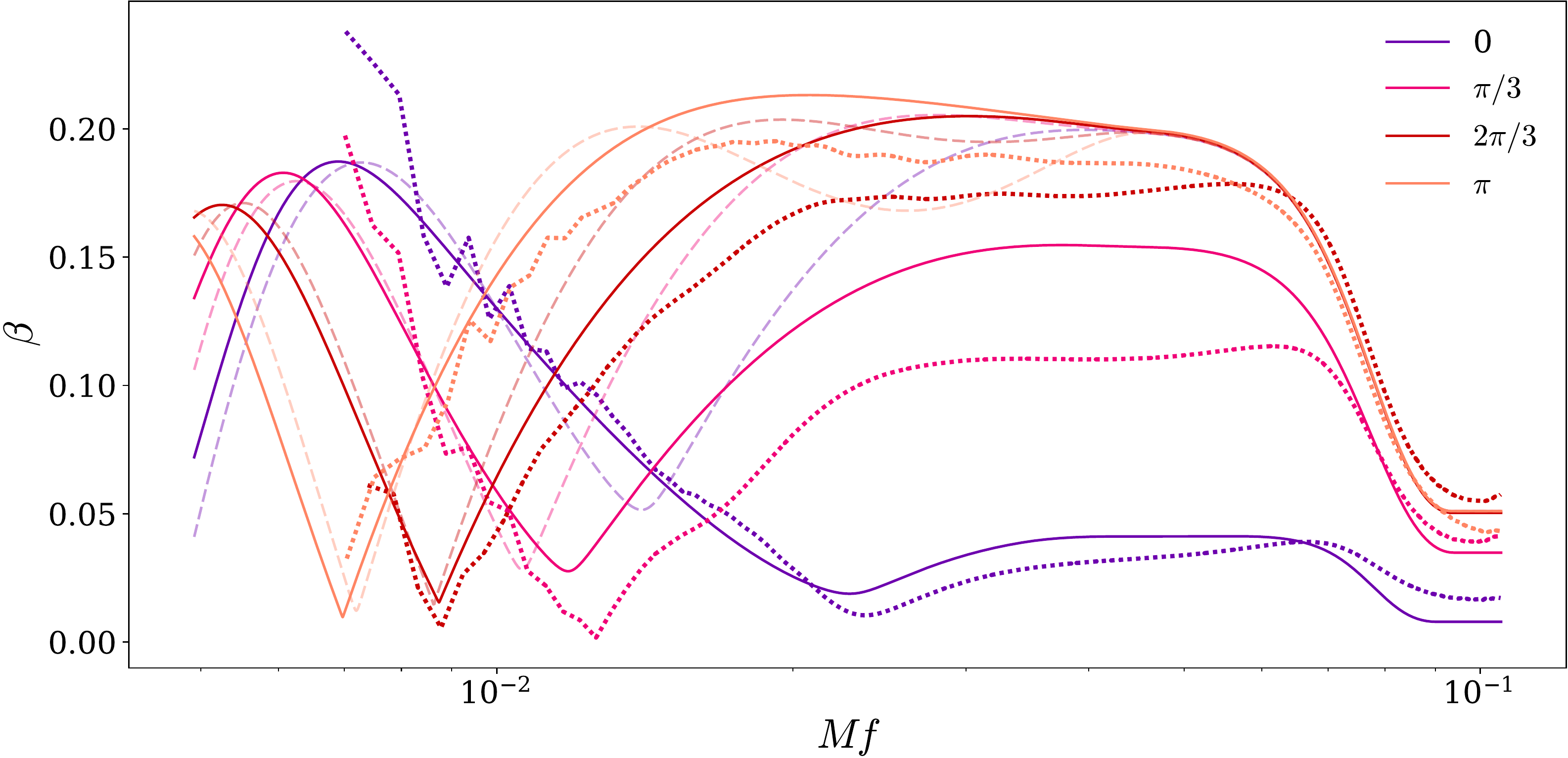} 
   \caption{Comparison of the Euler angle  $\beta$ between BAM NR simulations (dotted lines), \textsc{PhenomXO4a} (dashed lines) and \textsc{ PhenomXPNR} (solid lines). We consider a set of 4 simulations with spin magnitudes $\chi_1 = 0.2$, $\chi_2 = 0.8$, both in-plane. In the simulation initial data, the spin on the primary is antiparallel to the separation $\mathbf{n} = \mathbf{r}_2 - \mathbf{r}_1$ between the black holes, while the spin at the secondary is rotated in-plane to have an angle of $\phi_1 = \{0, \pi/3, 2\pi/3, \pi\}$ with respect to the primary.}
   \label{fig: two spin beta}
\end{figure}

This mapping is particularly important at small mass ratios, where the spins can be of equivalent magnitude and can therefore combine constructively or destructively, leading to strong or weak precession effects depending on the relative in-plane spin angles.
For systems with unequal black holes masses, the spins rotate in the plane at different rates.
It is therefore essential to employ the spin components immediately prior to merger in this calculation. 
At larger mass ratios, this mapping is equivalent to one using an average through the two-spin oscillations, as was employed in the \textsc{PhenomXO4a} model.
This improved single spin mapping allows us to capture two-spin effects through merger and ringdown without detailed additional calibration, as is demonstrated in Fig.~\ref{fig: two spin beta}.

Note that we retain the single-spin mapping employed in \textsc{PhenomXO4a} as given in Eq. (19) of Ref.~\cite{Ghosh:2023mhc} for the anti-symmetric part of the waveform. We explain this choice in sec.~\ref{sec: asymmetry}.

\subsection{Co-precessing model}
\label{sec: cp model}

The co-precessing model employed in \textsc{PhenomXPNR} follows closely that presented previously in \textsc{PhenomXO4a}~\cite{Thompson:2023ase}.
We employ the \nr{}-calibrated aligned spin, higher multipole model \textsc{PhenomXHM}~\cite{Garcia-Quiros:2020qpx} as the base model in the co-precessing frame. 
We include modifications to the late inspiral, merger and ringdown parts of the waveform that describe the effect of precession on the underlying co-precessing waveform and how the approximation that the co-precessing waveform can be described by the equivalent aligned-spin waveform breaks down. For the dominant ($2,\pm2$) multipole these modifications are calibrated to \nr{} simulations. We additionally include antisymmetric contributions to the ($2,\pm2$) multipoles. For the higher-order multipoles, the only modification is through an ``effective ringdown frequency'', as detailed below. 

The modifications to the symmetric part of the co-precessing waveform are described in detail in Sec. III A. 1. of Ref.~\cite{Thompson:2023ase}. In brief, deviations $u_{k}$ are applied to each of the model coefficients $\lambda_{k}$ describing the dominant multipole modelled in \textsc{PhenomXAS}~\cite{Pratten:2020fqn}, giving the new value of the parameter
\begin{align}
   \lambda'_{k} = {}& \lambda_{k} + \chi\sin\left(\theta_\mathrm{LS}\right) u_k.
\end{align} 

The subdominant multipoles are not calibrated directly. Instead, we adjust only the ringdown frequency, since the waveform is modelled in the co-precessing frame rather than that in which the ringdown frequencies are calculated when using perturbation theory. The ringdown frequency in the co-precessing frame $\omega'_{\ell m}$ is given by~\cite{Hamilton:2023znn}
\begin{align}\label{eq:eff_RDfreq}
   \omega'_{\ell m} = {}& \omega_{\ell m} - m\left(1-|\cos\beta_\mathrm{RD}|\right)\left(\omega_{22}-\omega_{21}\right),
\end{align}
where $\omega_{\ell m}$ are the ringdown frequencies as calculated from perturbation theory and $\beta_\mathrm{RD}$ is the value to which the opening angle of the precession cone settles after merger.

Outside the calibration region we aim to ensure that the model varies smoothly across the parameter space and the behaviour is not pathological. We therefore employ the deviations to the aligned-spin waveform model parameters only up to $q=20$. Beyond this mass ratio, we employ the \textsc{PhenomXAS} waveform  with the ringdown frequency still adjusted to be the effective ringdown frequency given in Eq.~\ref{eq:eff_RDfreq}. The transition occurs smoothly over the range $q\in\left[10,20\right]$.

\subsection{Multipole asymmetries}
\label{sec: asymmetry}

For aligned-spin binaries, we have the reflection symmetry
\begin{align} 
   h_{\ell, m} = {}& (-1)^\ell h^*_{\ell, -m}
   \label{eq:mode_asym}
\end{align}
between the positive and negative $m$ multipoles $h_{\ell, m}$.
However, this symmetry is broken for systems with misaligned spins~\cite{Bruegmann:2007bri,Boyle:2014ioa}, for which individual modes do not satisfy Eq.~\ref{eq:mode_asym}, irrespective of the frame in which they are expressed.
Consequently, aligned-spin waveforms (which respect equatorial symmetry by construction) can only serve as an approximate description of \gw{} multipoles in the co-precessing frame~\cite{Kalaghatgi:2020gsq,Ramos-Buades:2023ehm}. The asymmetry between negative and positive $m$ modes results in the emission of linear momentum perpendicular to the plane of the binary which can lead to large recoil velocities of the final black hole~\cite{Bruegmann:2007bri}. 

The asymmetry can be quantified as an anti-symmetric combination of the $\pm m$ modes for each $(l,m)$. We include the dominant multipole asymmetry model of Ref.~\cite{Ghosh:2023mhc} in the coprecessing-frame and rotate to the inertial frame following Eqs. (18) and (19) of Ref.~\cite{Thompson:2023ase}. 
We give here a brief overview of this model for completeness and refer the reader to the aforementioned references for further details.

The complex, anti-symmetric \fd{} waveform, $A_a (f)e^ {i\phi_a(f)}$, is used to construct the dominant multipoles $h_{2,\pm2}(f)$ in the coprecessing-frame as
\begin{align}
   h_{2,2}(f) = {}& A_s (f)e^{-i\phi_s(f)} + A_a(f) e^{i\phi_a(f)}, \\
   h_{2,-2}(f) = {}& A_s (f)e^{i\phi_s(f)} - A_a(f) e^{-i\phi_a(f)}, 
\end{align}
where $A_s$ and $\phi_s$ represent the symmetric amplitude and phase respectively.
The model for the anti-symmetric amplitude is of the form
\begin{align} 
   A_a(f) = {}& \kappa(f) A_s(f),
\end{align}
where $\kappa(f)$ is given by a \pn{} estimate and an \nr{}-calibrated correction.
 The anti-symmetric phase has an interesting structure, where it evolves differently from the symmetric phase in the inspiral~\cite{Boyle:2014ioa} but quickly catches up with it near merger, as shown in Fig. 11 of Ref.~\cite{Ghosh:2023mhc}. Therefore, the anti-symmetric phase uses a piecewise construction 
\begin{align} 
   \phi_a(f) := {}& 
   \begin{cases}
   \frac{1}{2}\phi_s(f) + \alpha(f) & f < p f_\mathrm{RD} \\
   \phi_s(f) &  f \ge p f_\mathrm{RD}
   \end{cases},
\end{align}
where $p\in[0,1]$, to map the symmetric phase and the precession angle $\alpha(f)$ (described in Sec.~\ref{sec: angle model}) into $\phi_a(f)$. This feature becomes even more intriguing for higher multipoles and was first identified in Ref.~\cite{Estelles:2025}. The optimal choice of $p$ and the different phase offsets, to ensure smooth transition in the phase, are obtained by calibrating to single-spin \nr{} simulations and have been discussed in details in Sec. V of Ref.~\cite{Ghosh:2023mhc}  

As noted in Sec.~\ref{sec: single spin map}, we retain the single-spin mapping previously employed in \textsc{PhenomXO4a} for the antisymmetric part of the waveform. 
The in-plane spin dependence of the anti-symmetric amplitude has been absorbed in the PN estimate of $\kappa(f)$, as is evident from Eq.~(15) of Ref.~\cite{Ghosh:2023mhc}. The amplitude of the asymmetry depends only on the overall misaligned spin content and not on spin precession; therefore, employing the instantaneous value of $\chi_\perp$ would not be be particularly useful for the asymmetry model. Further, for near-equal-mass configurations, the spins can combine destructively to almost completely cancel at a given frequency and incorrectly suppress the asymmetry entirely. We therefore retain the previous mapping.

Comparisons of out-of-plane recoil velocities with the \textsc{NRSur7dq4} model revealed that the initial implementation of the asymmetry model in \textsc{PhenomXO4a} neglected to account for the angle between the in-plane spin and separation vector at the reference frequency while constructing the anti-symmetric phase. Fixing this issue led to excellent agreement in recoil velocity estimates between the two models (cf. Fig. 4 and discussion in Sec. III C in Ref.~\cite{Mielke:2024kya}).

\subsection{Final spin}

The ringdown part of the model relies on the ringdown frequencies and damping times of the \qnm{} present in the waveform, as calculated from perturbation theory. These can be predicted for a given final mass and spin of the binary. The mass and spin of the remnant black hole formed from the corresponding aligned-spin binary are calculated using fits to numerical relativity, as in \textsc{PhenomXPHM}~\cite{Jimenez-Forteza:2016oae}.
The mass of the remnant is largely unaffected by presence of in-plane spins, whereas the final spin of the remnant requires the addition of the in-plane component. Since the in-plane spins are conserved to leading order throughout the evolution of the binary, we calculate the final spin as
\begin{align} 
   \chi_\mathrm{f} = {}& \mathrm{sgn}\left({\cos\beta_\mathrm{RD}}\right)
   \sqrt{\left(\left(\frac{m_1}{M}\right)^2\chi_\perp\right)^2 + \left(\chi^\parallel_\mathrm{f}\right)^2}
   \label{eq:final_spin}
\end{align}
where $\chi_\perp$ is calculated using Eq.~\ref{eqn: chi perp}, using the reference spins.
$\chi^\parallel_\mathrm{f}$ is the value of the final spin predicted for the corresponding aligned-spin binary as given by Ref.~\cite{Jimenez-Forteza:2016oae}.
Note that Eq.~\ref{eq:final_spin} differs from previous methods of estimating the final spin in the method used to determine the region of the parameter space in which the final spin changes sign.
Previously, $\mathrm{sgn}\big(\chi^\parallel_\mathrm{f}\big)$ was used; assuming that the direction of the final spin is consistent with the mapping to an aligned-spin system.
Here, we instead take it to be given by $\mathrm{sgn}\left({\cos\beta_\mathrm{RD}}\right)$, which is instead predicated on the size of the opening angle of the precession cone during ringdown~\cite{Hamilton:2023znn}.

\begin{figure}[t] 
   \centering
   \includegraphics[width=0.48\textwidth]{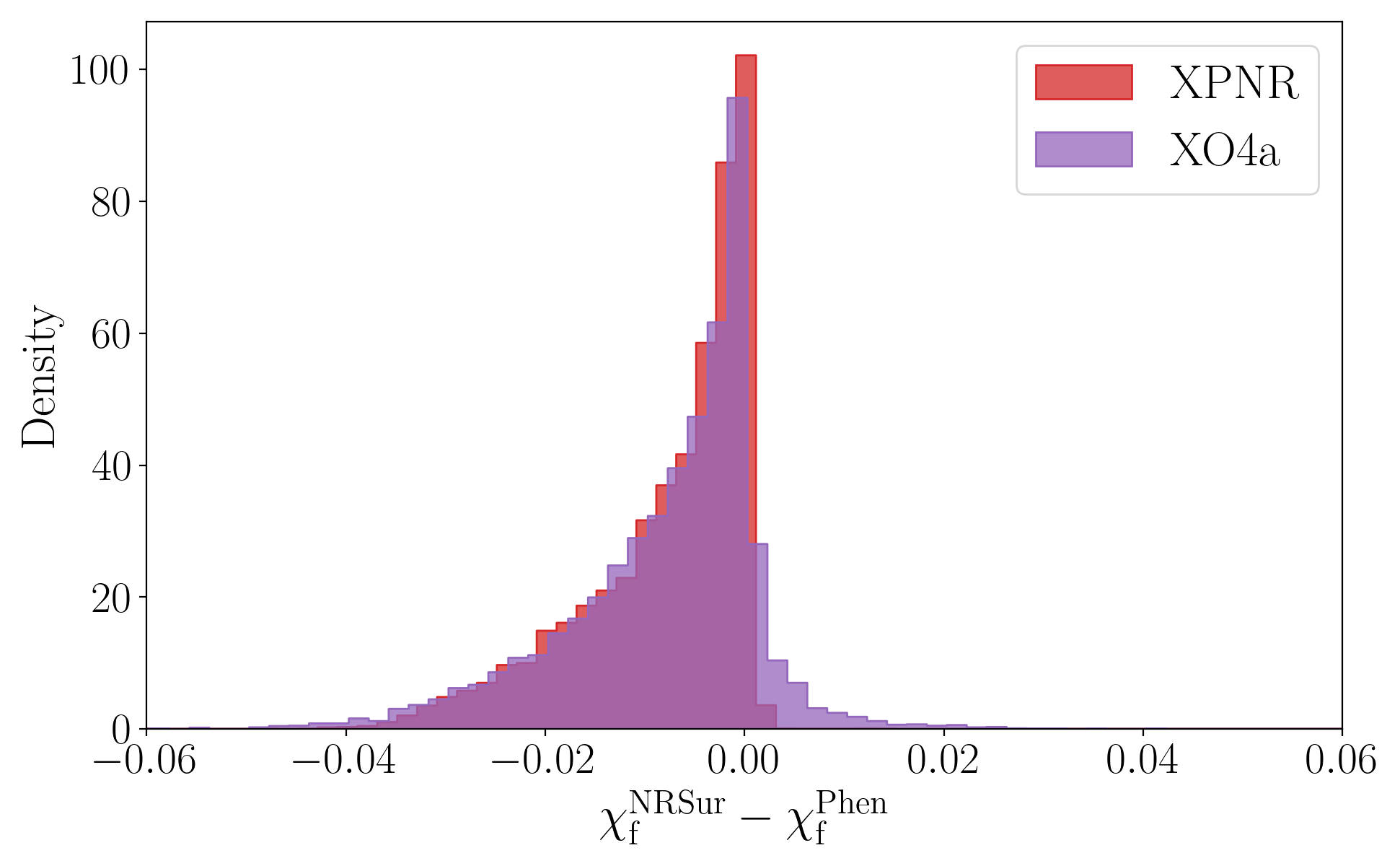} 
   \caption{Signed error on the remnant spin magnitudes predicted by \textsc{IMRPhenomXPNR} (red) and \textsc{IMRPhenomXO4a} (purple) with respect to the surrogate model \textsc{NRSur7dq4Remnant}. The errors have been evaluated on a set of 10000 randomly drawn precessing \bbh{s}. The final spin prescription employed by \textsc{IMRPhenomXPNR} yields slightly improved results, which are especially visible in the right tail of the error distribution.}
   \label{fig:final_spin}
\end{figure}

The choice to employ the sum of the in-plane spin components to calculate $\chi_\perp$ (as in Eq.~\ref{eqn: chi perp}) in the computation of the final spin magnitude yields a slightly more accurate estimate of the remnant spin than employing the precession-averaged $\chi_\mathrm{p}$, as done previously in \textsc{PhenomXO4a}. 
This is illustrated in Fig.~\ref{fig:final_spin}, where we compare the predictions of the surrogate model \textsc{NRSur7dq4Remnant} with the approximations employed in \textsc{PhenomXO4a} and \textsc{PhenomXPNR} for a sample of 10000 BBH configurations, with $q\in[1,5]$, $m_1\in[20,200]M_\odot$, spins isotropically distributed, and dimensionless spin magnitudes $\chi_{1,2}\leq 0.9$. For each configuration, the reference epoch of \textsc{NRSur7dq4Remnant} is chosen in accordance with the spin reference frequency of the phenomenological models.
The final spin predicted by \textsc{PhenomXPNR} (\textsc{PhenomXO4a}) has a median relative error of 0.8\% (1\%) with respect to the NR surrogate model. 

\subsection{Angle model}
\label{sec: angle model}

To transform the co-precessing model into a complete model of a precessing waveform, we require a model for the precession angles $\{\alpha,\beta,\gamma\}$. The model presented here for $\alpha$ and $\beta$ employs a \pn{} prescription to describe the evolution of the precession dynamics through the inspiral and a phenomenological description calibrated to \nr{} in the merger and ringdown. 
These two regions are then smoothly connected.
The remaining precession angle $\gamma$ is calculated numerically using the minimal rotation condition~\cite{Boyle:2011gg}.

For the precession angles $\alpha$ and $\beta$, we use the \st prescription during the inspiral, following the method described in~\cite{Colleoni:2024knd}: after numerically evolving the PN orbit-averaged spin-precession equations in the \td{}, one obtains $\alpha$ and $\beta$ as a function of frequency by means of the stationary phase approximation. In the current implementation of the model, we rely by default on the 3.5 PN \textsc{SpinTaylorT4} phasing, with 3 PN spin effects in the phasing and spin-precession equations. In the \st model, the precessional motion is tracked by the time-evolution of the Newtonian orbital angular momentum: hence, we rescale $\beta$ to ensure these angles describe the precessional motion of the \oed{} of the \gw{s} rather than of the plane of the binary. 
The rate of precession of the \oed{} during inspiral was shown in Ref.~\cite{Hamilton:2021pkf} to be consistent with the rate of precession of the Newtonian orbital angular momentum; consequently no further modification of the \st prescription is required for $\alpha$.

The merger-ringdown model for the angles is described in detail in Refs.~\cite{Thompson:2023ase} and~\cite{Hamilton:2021pkf}. 
The functional forms of $\alpha$ and $\beta$ are 
\begin{align} 
   \alpha(f) - \langle \alpha(f)\rangle = {}& 
   -\left( \frac{A_1}{f} + \frac{A_2\sqrt{A_3}}{A_3+\left(f-A_4\right)^2} \right), \\
   \beta(f) - \langle\beta(f)\rangle = {}&
   \beta(f) - B_0 = 
   \frac{B_1+B_2 f + B_3 f^2}{1+B_4\left(f+B_5\right)^2} \label{eqn: beta ansatz},
\end{align}
where $A_i$, $B_i$ are functions of $q$, $\chi$ and $\theta_\textrm{LS}$ given in Eqs. (25)-(35) of Ref.~\cite{Thompson:2023ase} and $\langle\rangle$ indicates the mean value. The coefficients in these ansatz are calibrated to a data set of 80 single-spin precessing BAM simulations~\cite{Hamilton:2023qkv}.
We employ the improved spin-mapping detailed in Sec.~\ref{sec: single spin map} in order to evaluate these coefficients for two-spin systems.

Outside the calibration region, the model for the merger-ringdown angles transitions smoothly to an analytic extension of the \st angles. To do this we employ the windowing function described in Sec. IV. C of Ref.~\cite{Thompson:2023ase}.
The analytic continuation of $\alpha$ is given by
\begin{align}
   \alpha(f) = {}& -\left(\varepsilon_1 + \frac{\varepsilon_2}{f^2} + \frac{\varepsilon_3}{f^4} \right),
\end{align}
where the coefficients $\varepsilon_i$ are determined by requiring that the ansatz match the \st value of $\alpha$ at the two frequency points $0.97 f^\mathrm{PN}_\mathrm{max}$ and $0.99 f^\mathrm{PN}_\mathrm{max}$ and its derivative at the latter point. $f^\mathrm{PN}_\mathrm{max}$ is the maximum frequency to which the \st integration can be performed.
The analytic continuation of $\beta$ is given by
\begin{align} 
   \beta(f) = {}& 
   \frac{e^{-\kappa f}}{f^2}\left( \xi_1 + \frac{\xi_2}{f} + \frac{\xi_3}{f^2} \right) 
   + \beta_\mathrm{RD},
\end{align}
where, as for $\alpha$, the coefficients $\xi_i$ are found from the requirement that the ansatz match the \st value at two frequency points (here $0.97 f^\mathrm{PN}_\mathrm{max}$ and $0.98 f^\mathrm{PN}_\mathrm{max}$) and it's derivative at the latter point. This ansatz ensures a smooth fall-off or rise to the value of $\beta_\mathrm{RD}$. Previously, in the \textsc{PhenomXPHM-SpinTaylor} model this value was given by 0 or $\pi$. 
However, we now use the fit to \nr{} for the ringdown value of $\beta$ given by Eq. (44) of Ref.~\cite{Thompson:2023ase}. The rate of decay is given by $\kappa = 2\pi\Delta f_\mathrm{damp}$, where $\Delta f_\mathrm{damp} = f^\mathrm{damp}_{\ell m} - f^\mathrm{damp}_{22}$ and $f^\mathrm{damp}_{\ell m}$ is the damping frequency of a given \qnm{} for a black hole with a given final mass and spin.

Inside the calibration region, where the calibrated model for the precession angles is employed, we ensure $C^1$ continuity by employing an intermediate region that matches the value and gradient of both the inspiral and merger-ringdown pieces at the connection frequencies.

For $\alpha$, the connection is done as described in Sec. VIII A of Ref.~\cite{Hamilton:2021pkf}.
We use an interpolating function of the form
\begin{align} 
   \alpha(f) = {}& a_0 f^2 + a_1 f + a_2 + \frac{a_3}{f},
\end{align}
where the coefficients $a_i$ are determined by requiring the interpolating function to be equal to the \st value and gradient at the start of the intermediate region and to the merger-ringdown ansatz value  and gradient at the end of the intermediate region.

The \st angles do not always evaluate up to the frequency required to connect to the merger-ringdown model for $\beta$, which is given by Eq.~\ref{eqn: beta ansatz}). We therefore introduce a connection function
\begin{align}
   \beta(f) = {}& b_0 + b_1 f + b_2 f^2 + b_3 f^3,
\end{align}
where the coefficients $b_i$ are determined by requiring that their value and gradient is given by that of the \pn{} $\beta$ at the end of the integration of the angles and by those of Eq.~\ref{eqn: beta ansatz} at the frequency at which the merger-ringdown ansatz is attached. We then treat this combined inspiral $\beta$ as described in Sec. VIII B of Ref.~\cite{Hamilton:2021pkf} and summarised below.

\begin{figure*}[t]
   \centering
   \includegraphics[width=\textwidth]{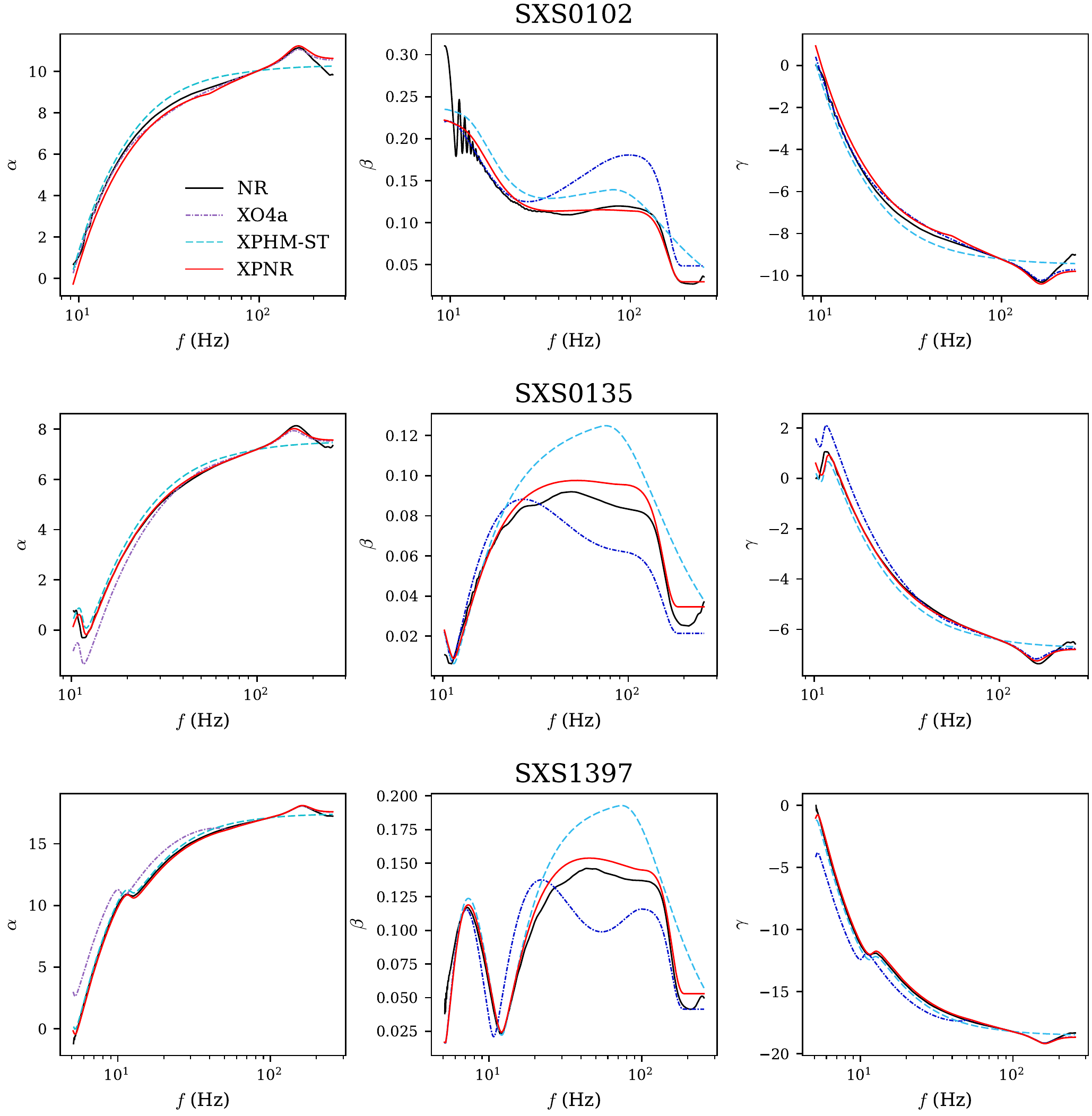} 
   \caption{Comparison of the new \textsc{IMRPhenomXPNR} model angles with those for a selection of \nr{} waveforms. We also compare the angles used in the previous models \textsc{IMRPhenomXPHM-SpinTaylor} and \textsc{IMRPhenomXO4a}. We consider three \nr{} cases at close to equal mass taken from the SXS catalogue~\cite{Boyle:2019kee}. Their properties are detailed in Tab.~\ref{tab: angle SXS cases}.}
   \label{fig: NR angle comparison}
\end{figure*}

First, we taper the oscillations seen in the \pn{} prescription for the angles during the inspiral as they are not present in the late inspiral to merger regions of signals from \nr{} simulations.
Further, it ensures a smooth connection with the merger-ringdown ansatz.
The tapered expression is then rescaled in order to find the value of $\beta$ describing the precession cone given by the \oed{} of the \gw{} signal, rather than that describing the evolution of the orbital angular momentum, as is returned by the \st evolution of the angles.
Finally, we apply an additional rescaling by multiplying through by an expression of the form 
\begin{align}
   k(f) = {}& 1 + k_1 f + k_2 f^2,
\end{align}
that leaves $\beta$ invariant at low frequencies but ensures the value of $\beta$ and it's derivative match at the connection frequency.

We transition smoothly beyond the calibration region over the range $q\in\{8.5,12\}$ and $\chi\in\{0.85, 1.2\}$.
To avoid the inclusion of pathologies in the fits outside the calibration region, we fix the values of the various coefficients $C_i\in\{A_i,B_i\}$ so that
\begin{align} 
   C_i\left(\eta < \eta_b; \chi > \chi_b \right) = {}& C_i\left(\eta = \eta_b; \chi = \chi_b \right), 
\end{align}
where $\eta_b = 0.09$ and $\chi_b = 0.8 - 0.2\exp\left[-((q-6)/1.5)^8\right]$.

\subsection{Extension to higher-order multipoles}
\label{sec: hm}

In the frequency domain, we cannot simply transform the higher-order multipoles with the set of precession angles appropriate for the $(2,\pm2)$ multipoles. 
Instead, we follow the prescription descbribed in Ref.~\cite{Khan:2018fmp} and further developed in Ref.~\cite{Thompson:2023ase} and use a set of angles applicable for each set of multipoles with a given $\ell$. 
The value of the higher multiple angles $\alpha$ and $\beta$ are obtained by a frequency rescaling of the relevant angle model for the $\ell=2$ multipoles. This operation mirrors the rescaling that is applied to the inspiral phasing of the subdominant modes to obtain their \fd{} evolution in terms of that of the $(2,\pm2)$ mode and rests on the same assumptions. 
At low frequencies, we employ the \pn{} frequency rescaling
\begin{align} 
   f \rightarrow {}& \frac{2f}{m},
\end{align}
while at high frequencies, we use the difference between the ringdown frequencies $f_{\ell m}^\textrm{RD}$ so
\begin{align} 
   f \rightarrow {}& f - \left(f_{\ell m}^\textrm{RD} - f_{22}^\textrm{RD}\right).
\end{align}
The final angle $\gamma$ is then obtained for each new pair of angles $\{\alpha,\beta\}$ via the minimal rotation condition, as for the $\ell=2$ angles.

\section{Model Performance and Parameter Estimation}
\label{sec:performance}

Combining the key features of the \textsc{PhenomXPHM-SpinTaylor} model with those of \textsc{PhenomXO4}a results in an improved description of the precessional dynamics of a binary system, as has been detailed above. 
In this section we demonstrate this improved description and explore the impact on the traditional measure of model accuracy --the mismatch-- and on the model's ability to accurately infer the parameters of a binary from a detected signal.
We employ the definitions of the mismatch $\mathfrak{M}$ and precessing mismatch $\mathfrak{M}_\mathrm{W}$ given in Sec. XI A of Ref.~\cite{Hamilton:2021pkf}, employing the power spectral density of advanced LIGO at design sensitivity~\cite{LSC:1800044-v5}.
Finally, we consider the impact of these developments on the computational performance of the model.

\subsection{Angle model accuracy}
\label{subsec:angle_accuracy}

\begin{table}[t]
   \centering
   \begin{tabular}{@{} lccc @{}} 
      \toprule
       & $q$ & $\mathbf{S}_1$ & $\mathbf{S}_2$ \\
      \midrule
      SXS0102 & 1.50 & (\! \phantom{-}0.50, \phantom{-}0.00, \phantom{-}0.00 \!) & (\! \phantom{-}0.50, \phantom{-}0.00, \phantom{-}0.00 \!) \\
      SXS0135 & 1.64 & (\!           -0.09, \phantom{-}0.06, \phantom{-}0.02 \!) & (\! \phantom{-}0.11,           -0.23,           -0.23 \!) \\
      SXS1397 & 1.56 & (\! \phantom{-}0.22, \phantom{-}0.09,           -0.18 \!) & (\!           -0.44,           -0.16, \phantom{-}0.11 \!) \\
      \bottomrule
   \end{tabular}
   \caption{Properties of \nr{} cases considered in Fig.~\ref{fig: NR angle comparison}.}
   \label{tab: angle SXS cases}
\end{table}

\begin{figure}[t] 
   \centering
   \includegraphics[width=0.49\textwidth]{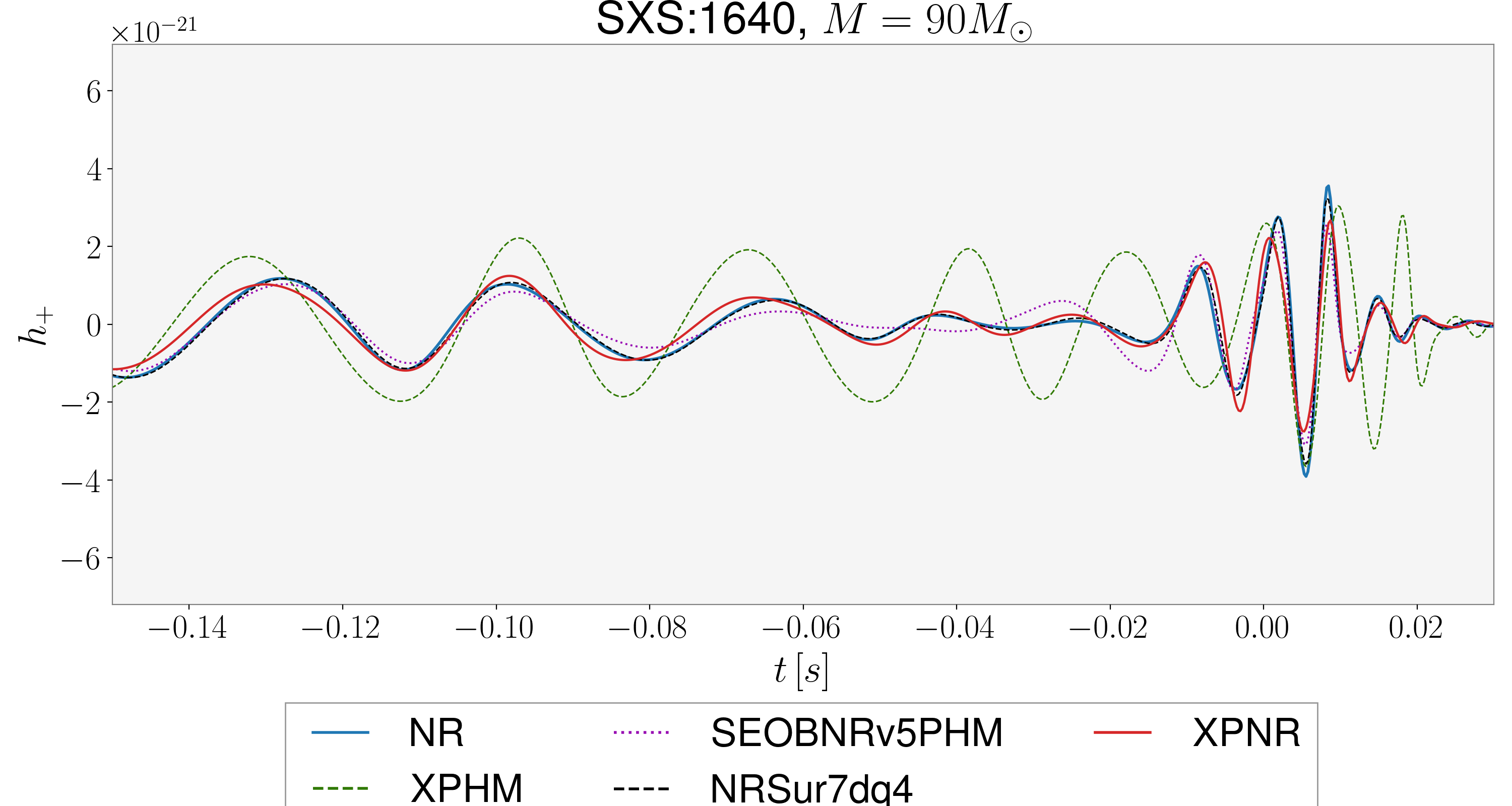} 
   \caption{\td{} comparisons between the NR simulation SXS:1640 and several time and \fd{} waveform models, for an illustrative binary with total mass of 90 $M_\odot$. The new model is plotted in red, and shows an excellent agreement with the underlying NR data (thick blue line).}
   \label{fig:td_comparison}
\end{figure}

Employing the \st angles improves the accuracy of the precessing waveform during inspiral, as has been demonstrated in Ref.~\cite{Colleoni:2024knd}.
In particular, the \st angles capture the phasing of two-spin oscillations with greater accuracy than the MSA angles previously employed in models such as PhenomXPHM-MSA and PhenomXO4a.
Further, since this approach involves the evolution of the precession dynamics, we are able to employ the values of the spins at merger in the single-spin mapping detailed in Sec.~\ref{sec: single spin map}, rather than using the spins at the reference frequency.
This ensures that the evaluation of the merger-ringdown ansatz, which relies on this mapping, is now significantly more accurate.
The greatest improvement is seen for systems at close to equal mass where the two-spin oscillations are strongest and the approximate two-spin mapping employed in \textsc{IMRPhenomXO4}a breaks down.

The improvement in the description of the precession dynamics from combining the \st angles with the \nr{}-informed merger-ringdown prescription is demonstrated in Fig.~\ref{fig: NR angle comparison}. 
Here we compare the angles extracted from NR simulations with those predicted by the three different models \textsc{PhenomXPHM-SpinTaylor}, \textsc{PhenomXO4}a and \textsc{PhenomXPNR} for three cases taken from the SXS catalogue~\cite{Boyle:2019kee}, details of which are given in Tab.~\ref{tab: angle SXS cases}.
It can clearly be seen that employing the \st angles, as in \textsc{PhenomXPHM-SpinTaylor} and \textsc{PhenomXPNR}, results in a more accurate description of the oscillations from two-spin systems.
Likewise, the inclusion of the calibration to \nr{} in the merger-ringdown in \textsc{PhenomXO4}a and \textsc{PhenomXPNR} ensures we capture crucial features of the angles.
For example, we capture the Lorentzian dip in the precession angle $\alpha$ and the rapid fall-off in $\beta$ which settles down to a constant value. 
Crucially, the use of the \st evolution of the spins to evaluate the merger-ringdown ansatz ensures an improved performance in the estimation of the magnitude of $\beta$ at the start of the merger. 
This means that the opening angle of the precession cone, and consequently the relative amplitude of the subdominant multipoles relative to the dominant $\ell=m$ multipole will be correctly captured throughout the merger and ringdown.

\begin{figure*}[t] 
   \centering
   \includegraphics[width=\textwidth]{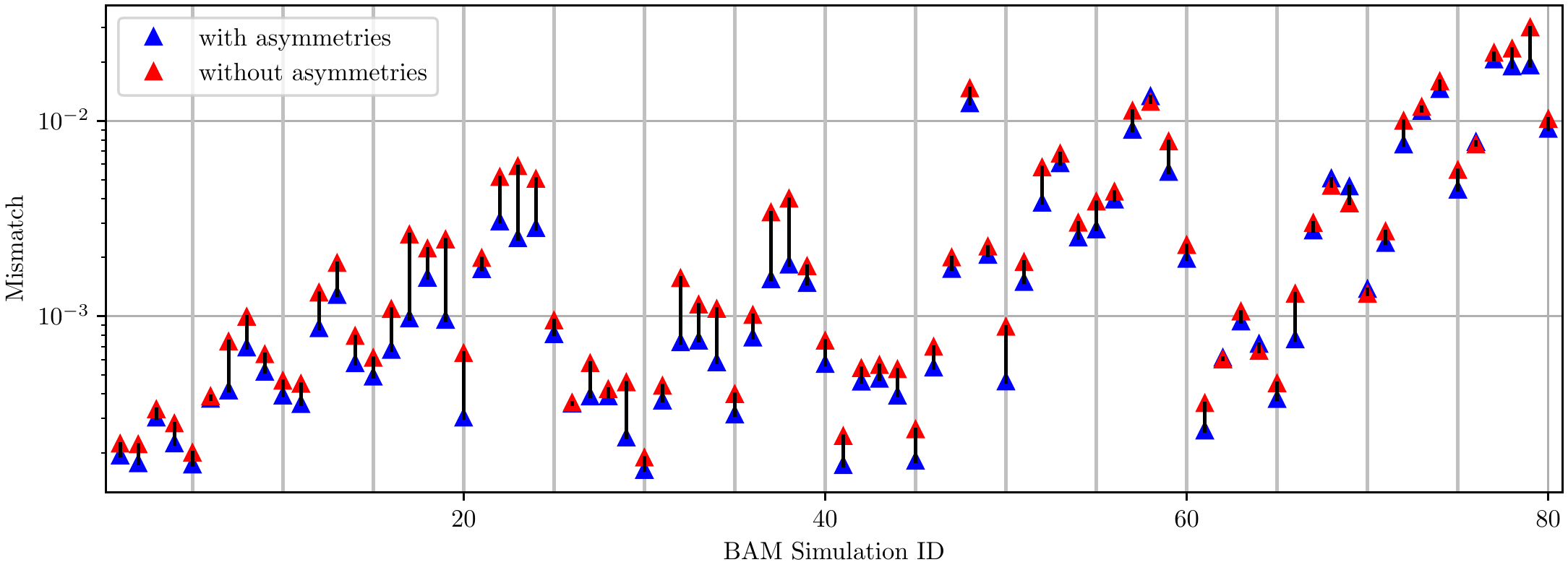} 
   \caption{Mismatch comparisons of \textsc{IMRPhenomXPNR} against BAM NR simulations with (\textit{blue}) and without (\textit{red}) asymmetries. The \textit{x}-axis shows the case identifiers for the NR simulations as used in the BAM catalog~\cite{Hamilton:2023qkv}.}
   \label{fig:asymmetry_matches}
\end{figure*}

This improvement in capturing the precession dynamics is translated to an overall improvement in the waveform.
In Fig.~\ref{fig:td_comparison}, we plot the cross polarization for the SXS simulation SXS:1640~\cite{Boyle:2019kee}, picking a total mass of 90$M_\odot$, with $\iota=\pi/8$ and reference frequency $f_{\mathrm{ref}}=15$ Hz. We superimpose the NR waveform with the prediction of several waveform models; for \fd{} waveforms, we first optimise over the reference phase and in-plane spins to find the best matching template. One can appreciate that \textsc{PhenomXPNR} shows an agreement with NR that is comparable to state-of-the-art \td{} models, greatly improving upon the original version of \textsc{PhenomXPHM}. 
Such reliable descriptions are essential for accurate measurements of the spin properties of high-mass binaries, such as GW190521~\cite{LIGOScientific:2020iuh}, where we see only the last few cycles of the signal. 

\subsection{Asymmetry model accuracy}

We demonstrate here the direct improvement to the model from the inclusion of mode asymmetries.
While the improvement in model accuracy due to the inclusion of dominant multipole asymmetry in the coprecessing frame was reported for \textsc{PhenomXO4a}~\cite{Thompson:2023ase}, the extent to which this improvement carries over when transforming into the inertial frame has not been investigated before. 
We compute the mismatch between the model and the 80 BAM \nr{} simulations~\cite{Hamilton:2023qkv} used in calibrating the model.
We consider only the $(2,\pm2)$ multipole in the co-precessing frame, since \textsc{PhenomXPNR} includes asymmetries only in the dominant multipole.
Furthermore, the inclusion of higher multipoles would make it harder to isolate the effects of the asymmetries.

The match is performed over the frequency range $f_\mathrm{min} = \mathrm{max}\left\{20, 1.35f_\mathrm{min}^\mathrm{NR}\right\}$Hz to $f_\mathrm{max} = 512$Hz, where $f_\mathrm{min}^\mathrm{NR}$ is the starting frequency of the relevant \nr{} waveform at a given total mass.
We considered systems at two different total masses: 60$M_\odot$ and 150$M_\odot$ and a range of 5 equally spaced inclinations between 0 and $\pi$ inclusive.

The results of this comparison are shown in Fig.~\ref{fig:asymmetry_matches}, where we plot the simulations by their identifier in the BAM catalog. 
For visual reference, the values shown in this figure are averaged over total mass and inclination.
The configurations shown here increase in mass ratio from left to right. Within each set of 20 simulations with the same mass ratio, the dimensionless spin magnitude increases.
We can see the expected parameter space trend of increasing mismatch with increasing mass ratio and spin magnitude as the precession effects become stronger and modelling errors have greater impact.
We demonstrate that the mismatches between \textsc{PhenomXPNR} and the BAM \nr{} simulations show noticeable improvement with the inclusion of mode asymmetries across the parameter space of calibration.  

\subsection{Mismatch comparisons}
\label{subsec:mismatches}

In order to assess the overall accuracy of the model, we computed the mismatch between the model and \textsc{NRSur7dq4} for $\sim4100$ binaries.
We use \textsc{NRSur7dq4} as a proxy for \nr{} simulations as it enables us to obtain a more uniform coverage of the parameter space and within its calibration is typically an order of magnitude more accurate than other models~\cite{Varma:2019csw}.
We therefore do not expect modelling errors in \textsc{NRSur7dq4} to dominate our assessment of our model performance.

We consider each system at a total mass of \{60, 90, 120, 150\}$M_\odot$ and at a range of six equally spaced inclinations between 0 and $\pi$ inclusive.
The intrinsic binary parameters are drawn from a uniform distribution over the calibration range of \textsc{NRSur7dq4}, sampling in the range of mass ratios $q\in\{0.25,1\}$ and dimensionless spin magnitudes $\chi_i\in\{0,0.8\}$.
The match is performed over the frequency range $f_\mathrm{min} = \mathrm{max}\left\{20, 1.35f_\mathrm{min}^\mathrm{NRSur}\right\}$Hz to $f_\mathrm{max} = 512$Hz, where $f_\mathrm{min}^\mathrm{NRSur}$ is the lowest starting frequency for \textsc{NRsur7dq4} for a given configuration at a given total mass.

In order to asses whether the model is sufficiently accurate that we expect to see unbiased parameter recovery for a binary at a given signal-to-noise ratio (SNR), we consider a conservative distinguishability criterion~\cite{Baird:2012cu}. 
For the measurement of a single parameter, we consider a 1-dimensional parameter space.
We therefore require a mismatch $\mathfrak{M} < 1.35/\rho^2$ to be confident that the model will not be biassed at 90\% confidence for a given SNR $\rho$.

\begin{figure}[t]
   \centering
   \includegraphics[width=0.49\textwidth]{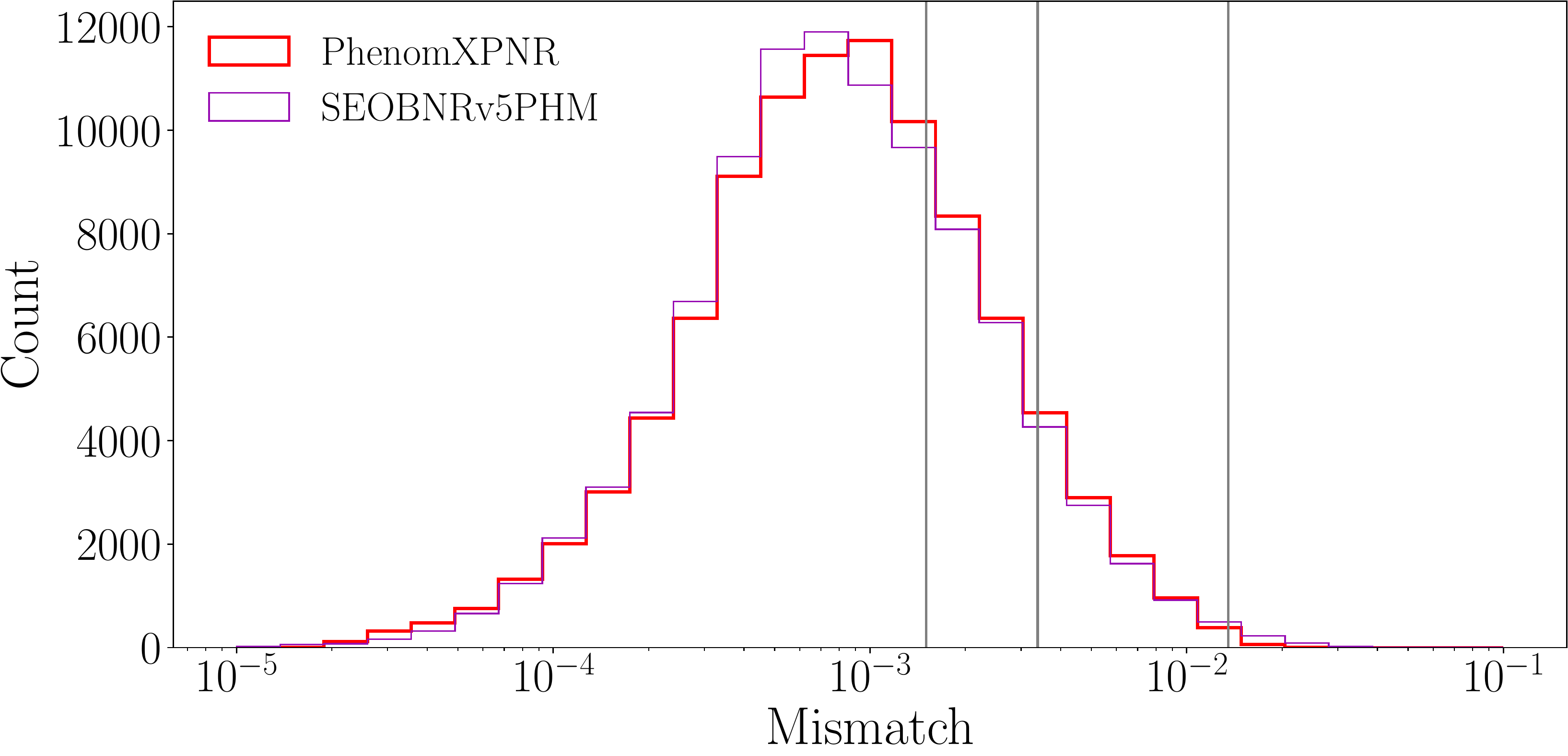} 
   \caption{Histogram showing the value of the mismatch value for $\sim4100$ binaries. We compute the mismatch between our chosen model and \textsc{NRSur7dq4}. We consider only the $\ell=2$ multipoles in the inertial frame. The grey lines give the mismatch threshold below which we are confident that the model should be unbiased at SNR 10, 20 and 30 (in order of decreasing mismatch) based on the distinguishability criterion.}
   \label{fig: ell2 mismatch histogram}
\end{figure}

When considering only the dominant quadrupole contribution to the signal (i.e. the $\ell=2$ multipoles) we find that the model performance is typically very good, as is shown in Fig.~\ref{fig: ell2 mismatch histogram}.
99.9, 90.9 and 71.8\% of cases satisfy the distinguishability criterion for signals at an SNR of 10, 20 and 30 respectively.
The model for the precession dynamics and the calibration of both the precession angles and the dominant multipole in the co-precessing frame is therefore performing well.

\begin{figure}[t]
   \centering
   \includegraphics[width=0.49\textwidth]{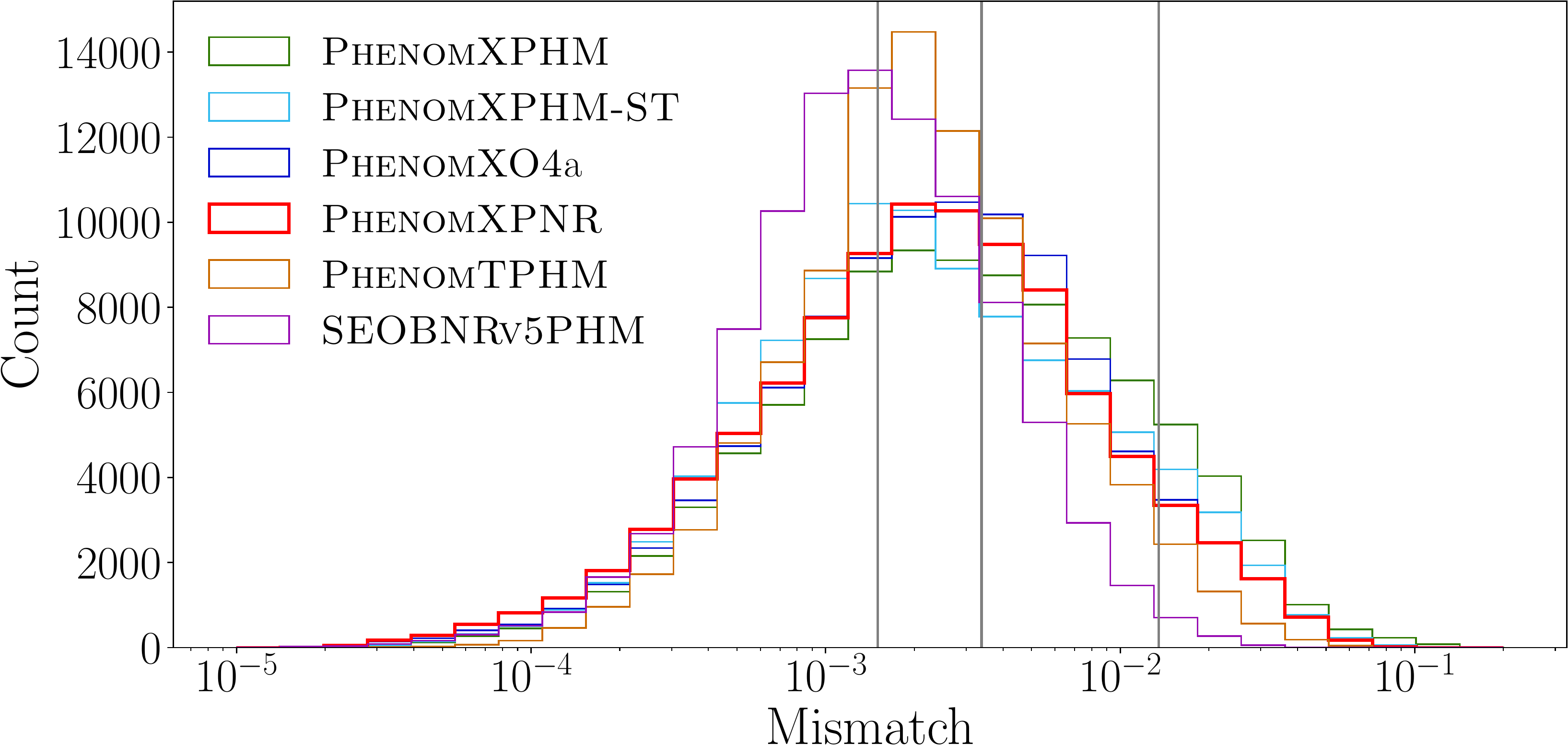} 
   \caption{Histogram showing the value of the mismatch value for $\sim4100$ binaries. We compute the mismatch between our chosen model and \textsc{NRSur7dq4}. We consider all available multipoles in the inertial frame. The grey lines give the mismatch threshold below which we are confident that the model should be unbiased at SNR 10, 20 and 30 (in order of decreasing mismatch) based on the distinguishability criterion.}
   \label{fig: complete mismatch histogram}
\end{figure}

However, when considering the complete signal with all multipoles included in the model, the performance degrades slightly, as can bee seen from Fig.~\ref{fig: complete mismatch histogram}, signalling we might need to reassess a number of our modelling assumptions for precessing systems in the frequency domain.
While the best performing cases in the low-mismatch tail are largely unaffected, the distribution now extends to much higher values of the mismatch ($\mathcal{O}\left(10^{-1}\right)$ as opposed to $\mathcal{O}\left(10^{-2}\right)$).
91.9, 62.8 and 37.9\% of cases satisfy the distinguishability criterion for signals at an SNR of 10, 20 and 30 respectively.
By contrast, for \textsc{SEOBNRv5PHM}, the best performing model shown in this figure, 99.1, 81.2 and 52.4\% of cases satisfy the distinguishability criterion for signals at an SNR of 10, 20 and 30 respectively.
We would therefore expect that, for signals above an SNR of 20, the parameter estimation recovery will be biassed for all models displayed in this figure across a reasonable fraction of the parameter space.
This is discussed further in Sec.~\ref{subsec:pe}.

\begin{figure*}[t]
   \centering
   \includegraphics[width=\textwidth]{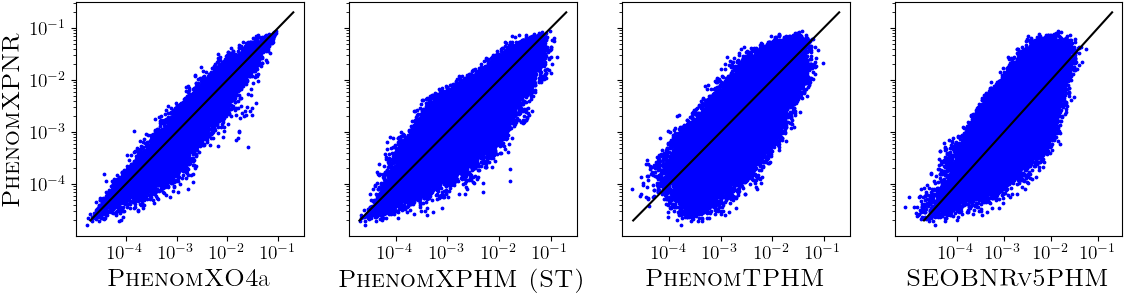} 
   \caption{Comparison of the performance of \textsc{PhenomXPNR} with other models for precessing black-hole-binaries. We show the mismatch for all $\sim4100$ binary configurations considered.}
   \label{fig: model comparison mismatches}
\end{figure*}

The relative performance of the difference models can be seen from Fig.~\ref{fig: model comparison mismatches}, which shows the results shown in Fig.~\ref{fig: complete mismatch histogram} as a direct model-by-model comparison for each binary configuration considered.
As the distribution of points is largely symmetric about the diagonal line in all panels, we can see that the performance of all models is approximately equivalent.
However, the slight over-preponderance of points below the diagonal line in the first three panels shows that on average \textsc{PhenomXPNR} shows a minor improvement the three other models \textsc{PhenomXO4}a, \textsc{PhenomXPHM-SpinTaylor} and \textsc{PhenomTPHM}.
This is expected from combining the \st evolution of the angles present in both \textsc{PhenomXPHM-SpinTaylor} and \textsc{PhenomTPHM}, which improves the model performance for lower mass binaries, with the calibration of the precession angles through merger and ringdown present in \textsc{PhenomXO4}a, which improves the model performance for higher mass binaries.
This minor improvement can be seen in mean of the distribution which goes from 0.00508 and 0.00485 for \textsc{PhenomXPHM-SpinTaylor} and \textsc{PhenomXO4}a respectively to 0.00468 for \textsc{PhenomXPNR}.
However, when comparing against \textsc{SEOBNRv5PHM}, we see that
\textsc{SEOBNRv5PHM} does not extend to quite such high values as \textsc{PhenomXPNR} and the other models.
This can be best explained by considering the performance of the model with varying inclination.

\begin{figure*}[htbp]
   \centering
   \includegraphics[width=\textwidth]{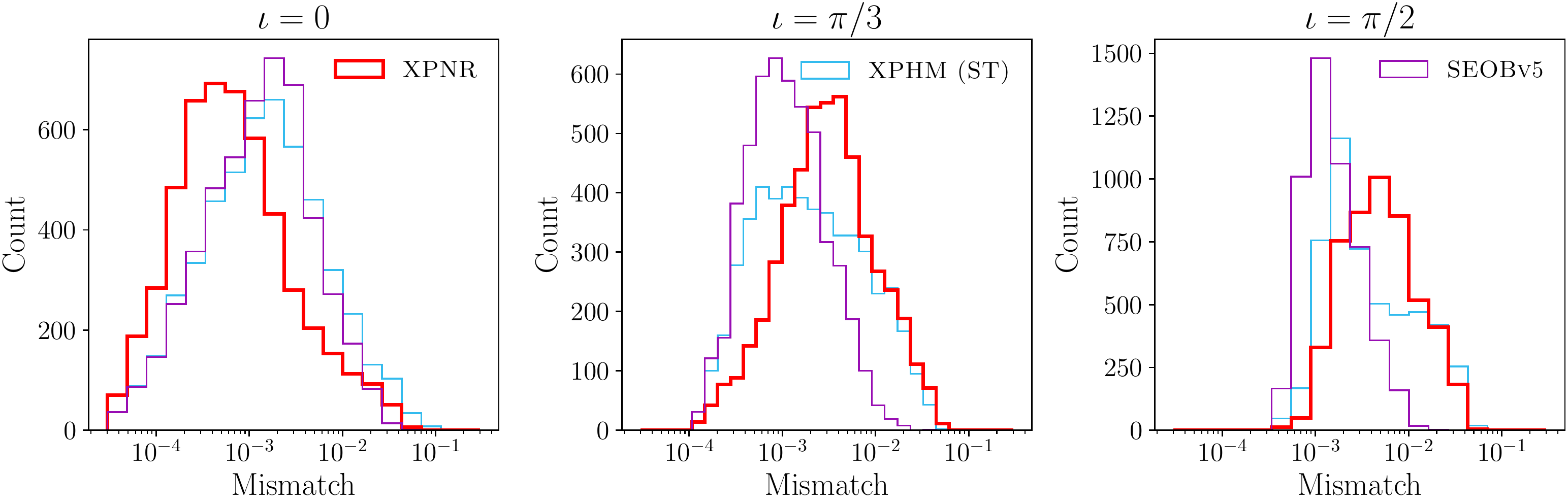} 
   \caption{We consider the variation in performance of three models, \textsc{PhenomXPNR}, \textsc{PhenomXPHM-SpinTaylor} and \textsc{SEOBNRv5PHM}, with inclination for high mass binaries (150$M_\odot$).} 
   \label{fig: inclination dependent mismatches}
\end{figure*}

The dependence of the mismatch on inclination for binaries with a total mass of 150$M_\odot$ is shown in Fig.~\ref{fig: inclination dependent mismatches}. 
It should be noted that for precessing binaries, the inclination $\iota$, as measured by the angle between the orbital angular momentum and the line of sight to the binary, will vary over the evolution of the binary as the orbital plane precesses. 
The values given here are therefore only true at the reference frequency. 
However, for the majority of cases considered here, the opening angle of the precession cone ($\beta$) will be sufficiently small as we are considering systems relatively close to equal mass.
Therefore, it is reasonable to assume that the cases labelled $\iota=0$ are close to face-on while those labelled $\iota=\pi/2$ are close to edge-on.

From Fig.~\ref{fig: inclination dependent mismatches} we can see that the improvements made to the model, and in particular the calibration of the co-precessing multipoles and merger-ringdown angles to \nr{}, results in comparatively low mismatches for systems close to face-on. 
However, this improvement is lost as we move to considering systems close to edge-on where the higher multipoles have a greater contribution. 
This is due to the approximations made when extending the precession angles to higher-order multipoles as detailed in Sec.~\ref{sec: hm}.
These approximations are not required when modelling precession in the \td{}, as is done for \textsc{SEOBNRv5PHM}.
It can clearly be seen that the performance of \textsc{SEOBNRv5PHM} is significantly less affected by the change in inclination than \textsc{PhenomXPNR} and \textsc{PhenomXPHM-SpinTaylor}.
Consequently the relative performance of the models shifts as the inclination angle increases, with \textsc{PhenomXPNR} the best performing model for ``face-on'' systems while \textsc{SEOBNRv5PHM} performs best for ``edge-on'' systems.

\begin{figure}[htbp]
   \centering
   \includegraphics[width=0.49\textwidth]{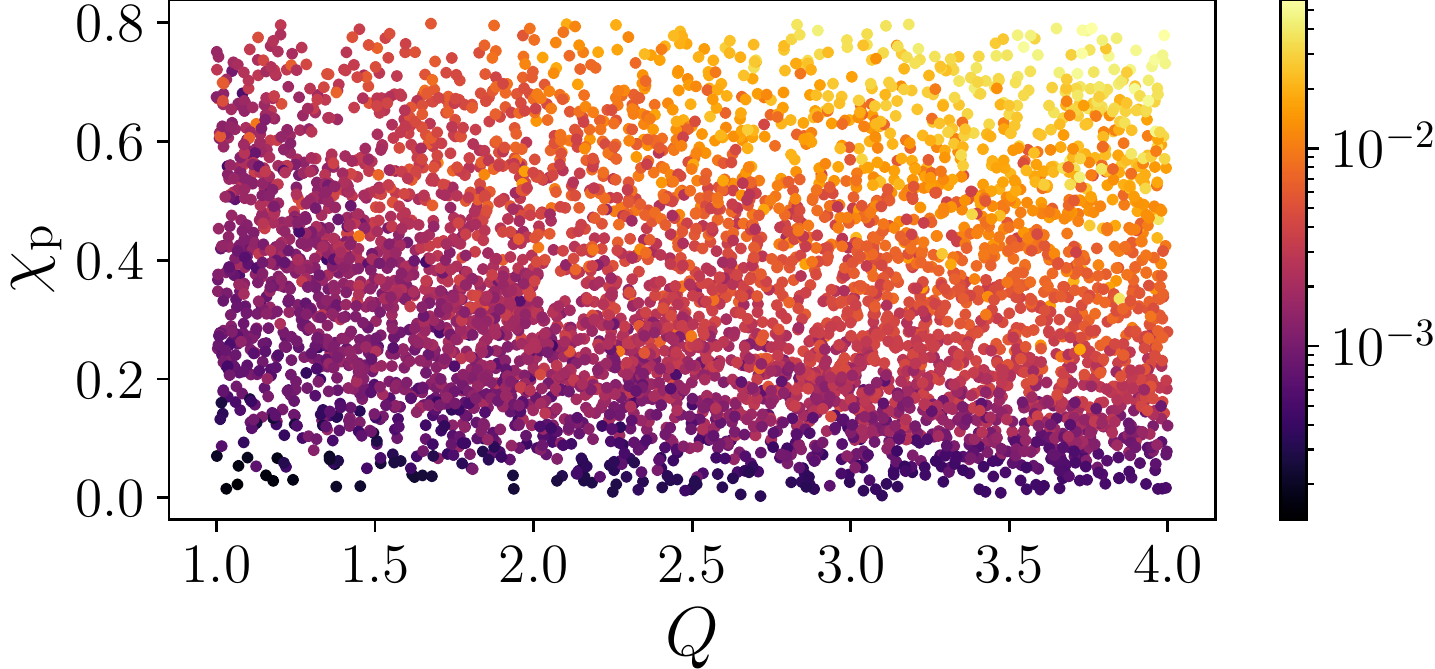} 
   \caption{Variation of the mismatch value averaged across all inclinations across parameter space for the model \textsc{PhenomXPNR}. Model performance is best for systems with low mass ratios and low precession.}
   \label{fig: mismatches parameter space dependence}
\end{figure}

This degradation in model performance can also be seen when considering the performance of the model across parameter space, as is shown in Fig.~\ref{fig: mismatches parameter space dependence}.
The model performs best for low-mass, low-precession signals with performance decreasing with increasing mass ratio and increasing value of $\chi_\mathrm{p}$.
This is in part due to the decreased fidelity of the model of the precession dynamics for a more highly precessing system and in part due to the increased contribution of the higher-order multipoles to the waveform of such a system at any inclination.

Overall, we see that the model \textsc{PhenomXPNR} is more accurate than either \textsc{PhenomXPHM-SpinTaylor} and \textsc{PhenomXO4a} and as such supersedes them as the most accurate \fd{} model of precessing quasi-circular \bbh{} systems currently available.
However, the \td{} model \textsc{SEOBNRv5PHM} continues to perform better by a factor of 2 on average due to approximations which are made in the \fd{} when modelling precessing systems which are not required in the \td{}.
This is particularly relevant for systems with higher mass ratios and at higher inclinations. 
For heavy binaries at lower mass ratios and close to face-on, \textsc{PhenomXPNR} is the most accurate model considered here.
The improved performance for low-inclination binaries is particularly pertinent as these are the systems we are most likely to observe with current detectors.

\subsection{Parameter Estimation Results}
\label{subsec:pe}

To examine \textsc{PhenomXPNR}'s ability to infer the properties of a binary black hole 
merger, we perform Bayesian inference on \gw{} data, $d$, comprised
of either a simulated or real \gw{} signal, $h$, and noise, $n$;
Bayesian inference is the process of estimating the \emph{posterior probability distribution},
which defines the probability of the binary having parameters $\boldsymbol{\lambda}$ given 
the \gw{} data and model $\mathbb{M}$.
The posterior is obtained through Bayes'  theorem,
\begin{equation}
    p(\boldsymbol{\lambda} | d, \mathbb{M}) = \frac{\mathcal{L}(d | \boldsymbol{\lambda},\mathbb{M})\, \Pi(\boldsymbol{\lambda} | \mathbb{M})}{\mathcal{Z}},
\end{equation}
where $\mathcal{L}(d | \boldsymbol{\lambda}, \mathbb{M})$ is the likelihood defined as the 
probability of the data given the binary parameters and model,
$\Pi(\boldsymbol{\lambda} | \mathbb{M})$ is the prior defined as the probability of the binary
parameters given the model, and
$\mathcal{Z} = \int{\mathcal{L}(d | \boldsymbol{\lambda}, \mathbb{M})\, \Pi(\boldsymbol{\lambda} | \mathbb{M}) d\boldsymbol{\lambda}}$ is the evidence. 
For a
quasi-circular binary black hole merger, $\boldsymbol{\lambda}$ is a 15-dimensional vector:
8 dimensions quantifying the mass, $m_{i}$, and spin, $\mathbf{S}_{i}$, of each black hole,
$i$, and 7 dimensions quantifying the source luminosity distance, sky location, merger time,
phase, polarization and inclination angle.

In this work, we estimate the posterior probability distribution through stochastic
sampling~\cite{Skilling:2006}. Specifically, we perform nested sampling~\cite{Skilling:2006} with 
{\textsc{Dynesty}}~\cite{Speagle:2020} via the {\textsc{bilby}} 
library~\cite{Ashton:2018jfp,Romero-Shaw:2020owr}. When analysing real \gw{} signals, we 
employ the same configuration and settings as those used by the LIGO--Virgo--KAGRA
collaboration~\cite{KAGRA:2023pio}. For all other analyses, we consistently used 1000 live 
points, the {\textsc{bilby}} implemented {\textsc{rwalk}} sampling algorithm, and a theoretical
power spectral density for Advanced LIGO's~\cite{LIGOScientific:2014pky} and Advanced Virgo's~\cite{VIRGO:2014yos} design
sensitivity~\cite{dcc:T2200043}.

Unless otherwise stated, when analysing simulated \gw{} signals we take advantage of the
zero-noise
limit, defined as $n=0$ for all times. 
A zero-noise analysis is the expected result
when averaging over many different Gaussian noise realisations.
By calculating and comparing the posterior distribution obtained for a range 
of different models in zero-noise, we can provide further insights into the accuracy of 
\textsc{PhenomXPNR}; waveform differences can be probed
directly and
posteriors compared to the true source properties. We note that although a typical
Bayesian analysis evaluates the model $O(10^{7})$ times,
we can only gain a limited insight into the performance of \textsc{PhenomXPNR} over the
vast 15-dimensional parameter space for a handful of injections. Also, although there have been 
recent efforts to explicitly incorporate the mismatch results in Fig.~\ref{fig: complete mismatch histogram} into Bayesian
inference techniques~\cite{Hoy:2024vpc}, we do not employ these techniques in this work.

\subsubsection{Simulated \gw{} signals}

First, we assess the performance of {\textsc{PhenomXPNR}} for \gw{} simulations of known parameters. We pay particular attention to the improvement of 
{\textsc{PhenomXPNR}} compared to its predecessor {\textsc{PhenomXPHM-SpinTaylor}} to highlight the enhancements that we have implemented as part of this work. 
We consider two binary black hole systems where the properties
differ from current population estimates~\cite{KAGRA:2021duu}. The reason is because we 
expect the differences between models to be amplified in this region of the parameter space.

We consider a mass ratio $Q = 4$ \nr{} simulation with a
non-zero spin. 
We inject CF\_54 provided by the
BAM catalogue~\cite{Hamilton:2023qkv} with component masses $m_{1} = 48\, M_{\odot}$ and
$m_{2} = 12\, M_{\odot}$, primary spin magnitude $a_{1} = 0.6$ tilted at angle
$\theta_{1} = 2\pi/3 \, \mathrm{rad}$ and a secondary spin of 0. 
The inclination angle was defined to be $\theta_{\mathrm{JN}} = 0.66\, \mathrm{rad}$, and
the SNR was chosen to be $\rho = 20$. 
This simulation was selected since we observed an order of magnitude better performance (in terms of mismatches) for {\textsc{PhenomXPNR}} compared to {\textsc{PhenomXPHM-SpinTaylor}} and {\textsc{PhenomXO4a}}. For this simulation we calculated mismatches: 0.008, 0.015, 0.016, 0.005, 0.004 for  {\textsc{PhenomXPNR}}, {\textsc{PhenomXPHM-SpinTaylor}}, {\textsc{PhenomXO4a}}, {\textsc{SEOBNRv5PHM}} and {\textsc{NRSur7dq4}} respectively. Based on these mismatches, we would naively expect {\textsc{PhenomXPNR}}, {\textsc{SEOBNRv5PHM}} and {\textsc{NRSur7dq4}} to perform optimally for this system.

\begin{figure}[t]
   \centering
   \includegraphics[width=0.48\textwidth]{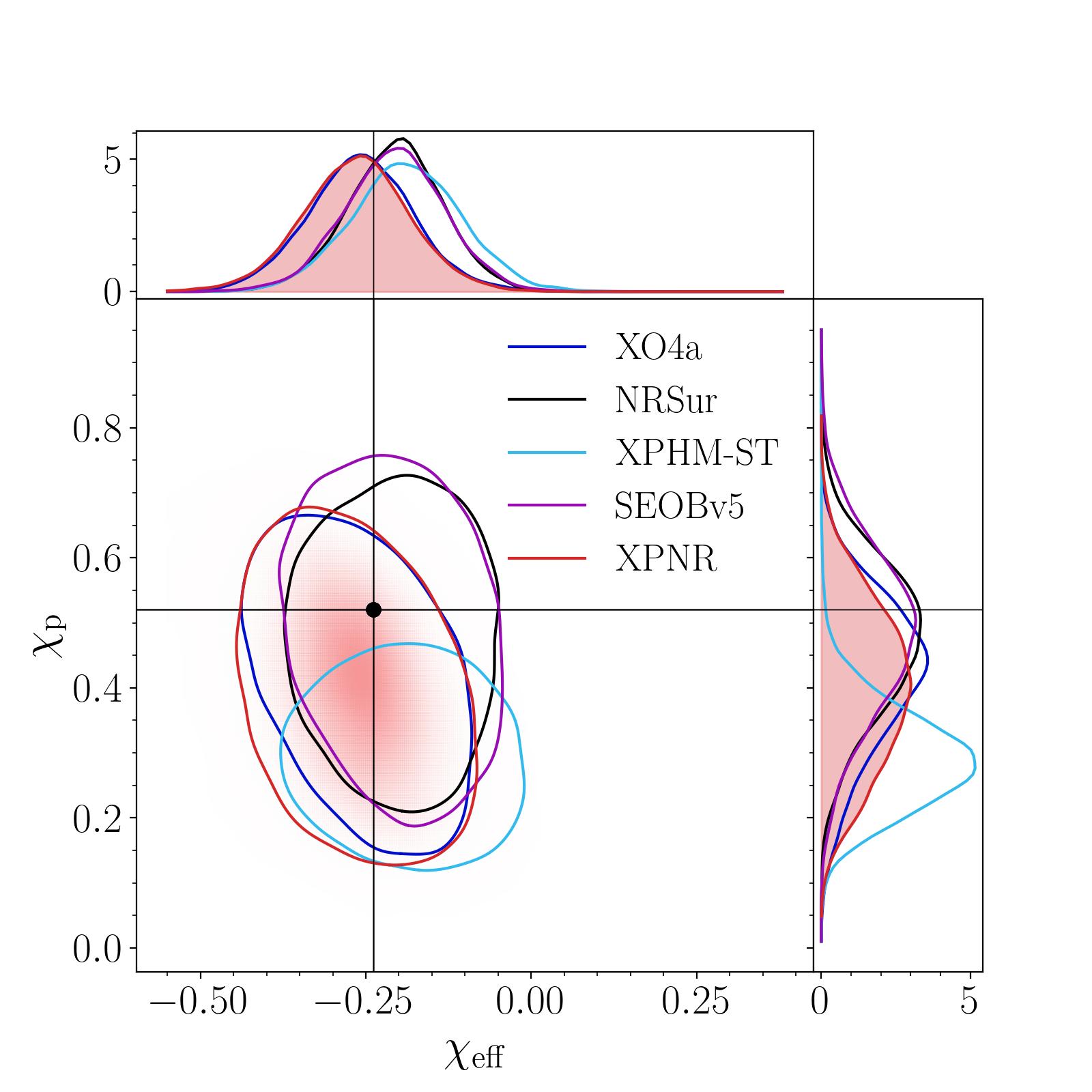} 
   \caption{Comparison between the two-dimensional marginalized posterior distributions for the effective parallel spin $\chi_{\mathrm{eff}}$ and effective perpendicular spin $\chi_{\mathrm{p}}$ when analysing the CF\kern-0.1em\_\kern+0.05em54 numerical relativity simulation~\cite{Hamilton:2023qkv}. The contours encase 90\% probability. The black crosshairs show the true values.}
   \label{fig:simulation_comparison_bam}
\end{figure}

In Fig.~\ref{fig:simulation_comparison_bam} we see that {\textsc{PhenomXPNR}} 
recovers the effective spins of the binary
within the 90\% confidence region, while {\textsc{PhenomXPHM-SpinTaylor}} misidentifies the
amount of precession in the binary: {\textsc{PhenomXPHM-SpinTaylor}} infers greater support
for a binary with spins aligned with the orbital angular momentum (smaller $\chi_{\mathrm{p}}$).
We understand that NR-tuned co-precessing dominant mode inherited from {\textsc{PhenomXO4a}} is the primary reason behind
{\textsc{PhenomXPNR}}'s improvement in the recovered effective precessing spin: when we
turn off the NR tuning for the co-precessing dominant mode we obtain a posterior distributions
that is in better agreement with the {\textsc{PhenomXPHM-SpinTaylor}} result. This is why
{\textsc{PhenomXPNR}} and {\textsc{PhenomXO4a}} recover comparable distributions
for this injection. 
When comparing Bayesian evidences, we see that
the data prefers {\textsc{PhenomXPNR}} over {\textsc{PhenomXPHM-SpinTaylor}} by a factor
of 3:1. We also analysed this injection with the
{\textsc{NRSur7dq4}} and {\textsc{SEOBNRv5PHM}} models. In general,
{\textsc{NRSur7dq4}} and {\textsc{SEOBNRv5PHM}} more accurately capture the true source
properties, with {\textsc{SEOBNRv5PHM}} closely resembling {\textsc{NRSur7dq4}} for this
analysis. We see that {\textsc{PhenomXPNR}} agrees better with
{\textsc{NRSur7dq4}} and {\textsc{SEOBNRv5PHM}} than {\textsc{PhenomXPHM-SpinTaylor}}.

Next, we inject a \gw{} signal produced by a high mass binary black hole system. 
We chose to analyse this region of the parameter space
since it enables a direct study of the merger/ringdown portion of the \gw{}
signal, allowing us to gauge the effects of the improved merger/ringdown prescription in
{\textsc{PhenomXPNR}} over {\textsc{PhenomXPHM-SpinTaylor}}.
We simulate a binary black hole merger with component masses $m_{1} = 100\, M_{\odot}$ and
$m_{2} = 50\, M_{\odot}$, primary and secondary spin magnitudes $a_{1} = a_{2} = 0.8$ with
tilts $\theta_{1} = 1.2\, \mathrm{rad}$ and $\theta_{2} = 1.5\, \mathrm{rad}$ for the primary
and secondary spin vector respectively. The inclination angle was defined to be $\theta_{\mathrm{JN}} = 0.45\, \mathrm{rad}$, and
the SNR was chosen to be 50. Although the SNR is
high, it remains less than the expected SNR at which GW150914~\cite{LIGOScientific:2016aoc,LIGOScientific:2016wkq}
would have been observed with O4a-like detector sensitivities ($\sim 70$)~\cite{Gaebel:2017zys}. We generate the \gw{} signal with {\textsc{NRSur7dq4}}.

\begin{figure}[t]
   \centering
   \includegraphics[width=0.48\textwidth]{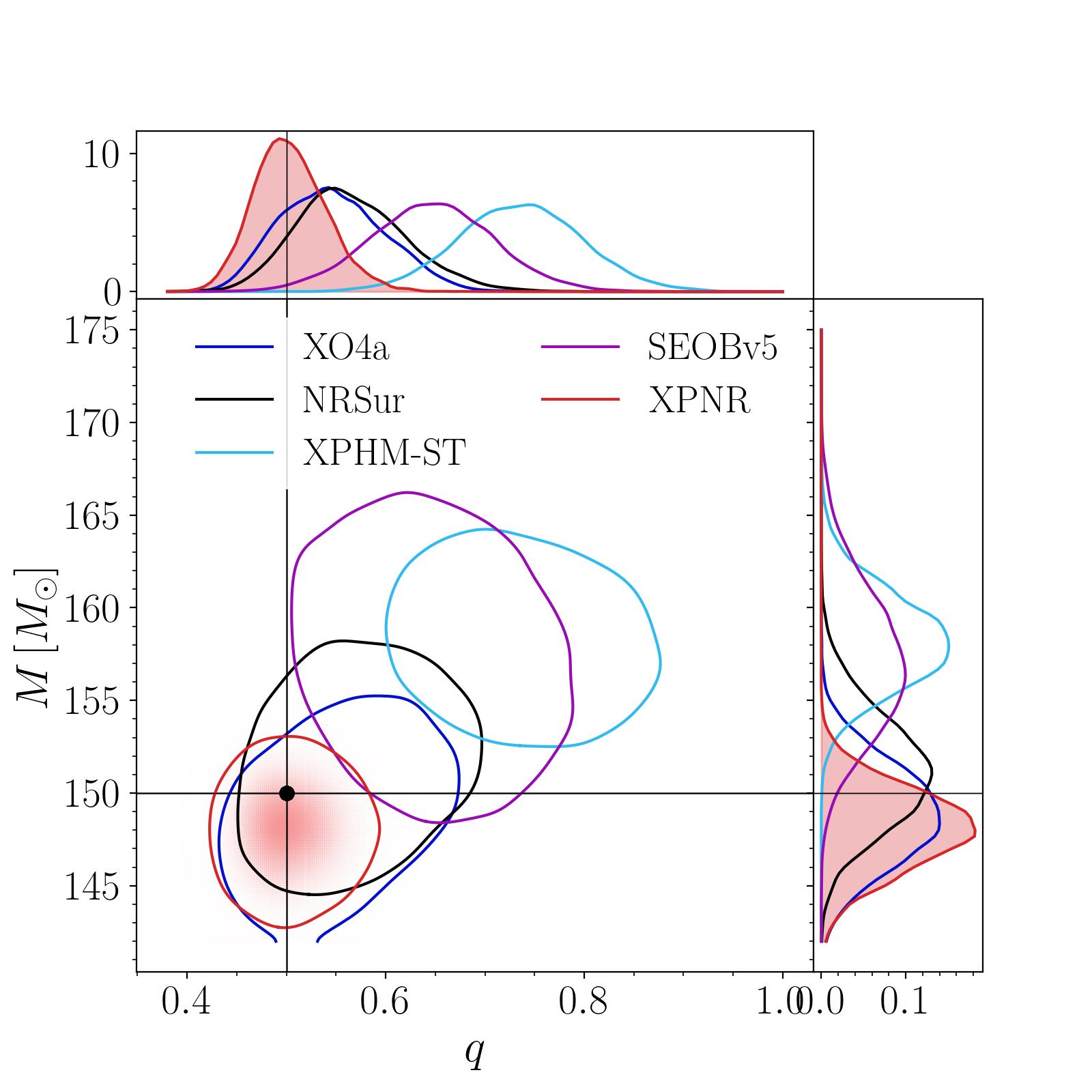}
   \caption{Comparison between the two-dimensional marginalized posterior distributions for the binary total mass $M$ and mass ratio $q$ when analysing an {\textsc{NRSur7dq4}} injection. The contours encase 90\% probability. The black cross hairs show the true values.}
   \label{fig:simulation_comparison_nrsur}
\end{figure}

In Fig.~\ref{fig:simulation_comparison_nrsur} we see that 
{\textsc{PhenomXPNR}} more accurately captures the binary's total mass and mass ratio;
all other models considered infer a higher total mass binary with more symmetric component
masses. Interestingly, {\textsc{PhenomXPNR}} and {\textsc{PhenomXPHM-SpinTaylor}}
recover disjoint distributions, with both {\textsc{PhenomXPHM-SpinTaylor}} and
{\textsc{SEOBNRv5PHM}} failing to recover the true source masses.
When comparing {\textsc{PhenomXPNR}} to
{\textsc{PhenomXO4a}}, we see that the additional improvements outlined in this paper lead directly to a tighter distribution around the true source masses.
Finally, {\textsc{PhenomXPNR}} more accurately captures the binary's true mass ratio and
total mass than {\sc{NRSur7dq4}} despite the simulation being made with {\sc{NRSur7dq4}}.
We highlight that although {\textsc{PhenomXPNR}} performs better in this two-dimensional
slice, the posterior is fifteen dimensional and overall {\sc{NRSur7dq4}} fits the data better than
{\textsc{PhenomXPNR}}. This can be seen by the larger Bayesian evidences for {\sc{NRSur7dq4}} compared to {\textsc{PhenomXPNR}}. 

The improved performance
of {\textsc{PhenomXPNR}} compared to the other models for this simulation correlates directly with the
 improved mismatch calculations. We find that {\textsc{PhenomXPNR}} obtains a mismatch of 0.003 when comparing to the injection at the same fiducial parameters. This compares to
 0.033 for {\textsc{PhenomXPHM-SpinTaylor}} and 0.026 for {\textsc{SEOBNRv5PHM}} -- both an order of magnitude larger than {\textsc{PhenomXPNR}} -- and 0.002 for {\text{PhenomXO4a}}. Assuming the commonly used distinguishability criterion from Ref.~\cite{Baird:2012cu} (although we note that this is a conservative metric~\cite{Thompson:2025hhc}), two waveforms are indistinguishable if their mismatch is below 0.003 for an SNR 50 GW signal (assuming 8 degrees of freedom). It is therefore unsurprising that {\textsc{PhenomXPNR}} and {\textsc{PhenomXO4a}} are the only models\footnote{Aside from {\textsc{NRSur7dq4}}, which has a mismatch of exactly 0, since this model was used to generate the injection.} to infer unbiased source estimates.
 
The observed waveform systematics in Fig.~\ref{fig:simulation_comparison_nrsur}
implies that care must be taken when analysing high mass binary black hole systems
with significant in-plane spin at large SNRs. Although astrophysically interesting, estimates
for the black hole masses may differ depending on the model selected when performing
inference. We note that this conclusion is dependent on the SNR of the injected signals.
The inferred 90\% confidence interval of each waveform model inversely scales with SNR:
a wider posterior will be obtained for a lower SNR injection. Therefore, if the SNR is reduced
we would expect better agreement between waveform models, with more models encasing the
injected value within the 90\% confidence interval. However, the maximum likelihood estimate, and posterior median should remain unchanged.

Although not shown here, we also verified the performance of {\textsc{PhenomXPNR}} for
100 randomly chosen \gw{} signals drawn from the prior distribution. We generated these
simulations with {\textsc{PhenomXPNR}},  injected the signals into different
instances of Gaussian noise for a network of two Advanced LIGO detectors~\cite{LIGOScientific:2014pky} and one Advanced Virgo detector~\cite{VIRGO:2014yos}, and assumed a duration of 8 seconds of data.
When comparing the inferred posterior distributions against the injected
values, we obtain the expectations from a Gaussian likelihood -- an approximately linear
relationship for each dimension.

\subsubsection{Observed \gw{} signals}

Next we consider the performance of {\textsc{PhenomXPNR}} for real \gw{} signals. We analyse two signals that show waveform systematics between published posterior distributions, and have claimed evidence for precession of the orbital plane~\cite{Hoy:2021dqg,Hoy:2024wkc}: GW190412~\cite{LIGOScientific:2020stg, Islam:2020reh} and GW200129~\cite{KAGRA:2021vkt,Hannam:2021pit}\footnote{We note that the evidence for precession in GW200129 has been argued against in previous work owing to noise artefacts at the time of GW200129~\cite{Payne:2022spz}, although see Ref.~\cite{Macas:2023wiw} for a counter argument which claims that the evidence for precession remains after improved data cleaning, and claims that the binary may be eccentric~\cite{Romero-Shaw:2022xko,Gupte:2024jfe,Planas:2025jny}.}. These signals are therefore natural choices to test {\textsc{PhenomXPNR}}'s improved precession modelling and to check it's consistency with other state-of-the-art models.
  
 In the left panel of Fig.~\ref{fig:event_comparison} we show the inferred effective aligned spin
 $\chi_{\mathrm{eff}}$ and mass ratio $q$ for GW190412. We compare the posterior obtained
 with {\textsc{PhenomXPNR}} to {\textsc{SEOBNRv5PHM}} and the published results by
 the LIGO--Virgo--KAGRA collaboration~\cite{LIGOScientific:2020stg}; the LVK
 published results with {\textsc{PhenomPv3HM}}~\cite{Khan:2018fmp}, labelled {\emph{Phenom PHM}}, and
 {\textsc{SEOBNRv4PHM}~\cite{Ossokine:2020kjp}, labelled {\emph{EOBNR PHM}}. When restricting attention to the
 results published by the LIGO--Virgo--KAGRA collaboration, we see possible waveform
 systematics with \emph{Phenom PHM} preferring more equal mass ratios and lower
 aligned spin than \emph{EOBNR PHM}.  We see that the observed
  systematics are no longer present amongst the latest generation of Phenom and SEOB 
 models: {\textsc{PhenomXPNR}} and {\textsc{SEOBNRv5PHM}} show good agreement with both inferring overlapping mass ratio and aligned spin probability densities. This corroborates the findings in Ref.~\cite{Colleoni:2020tgc}, which highlighted that even for the non-precessing parameter space, the {\textsc{PhenomX}} family of waveform models removes the observed systematics in GW190412 owing to the calibration to \nr{} of the higher order multipoles.

In the right panel of Fig.~\ref{fig:event_comparison} we show the inferred effective aligned
spin $\chi_{\mathrm{eff}}$ and effective precessing spin $\chi_{\mathrm{p}}$ for GW200129.
We see that {\textsc{PhenomXPNR}} and {\textsc{SEOBNRv5PHM}} agree well, but both
infer less precession than {\textsc{NRSur7dq4}} and {\textsc{PhenomXPHM-SpinTaylor}}. 

   \begin{figure*}[htbp]
   \centering
   \includegraphics[width=0.48\textwidth]{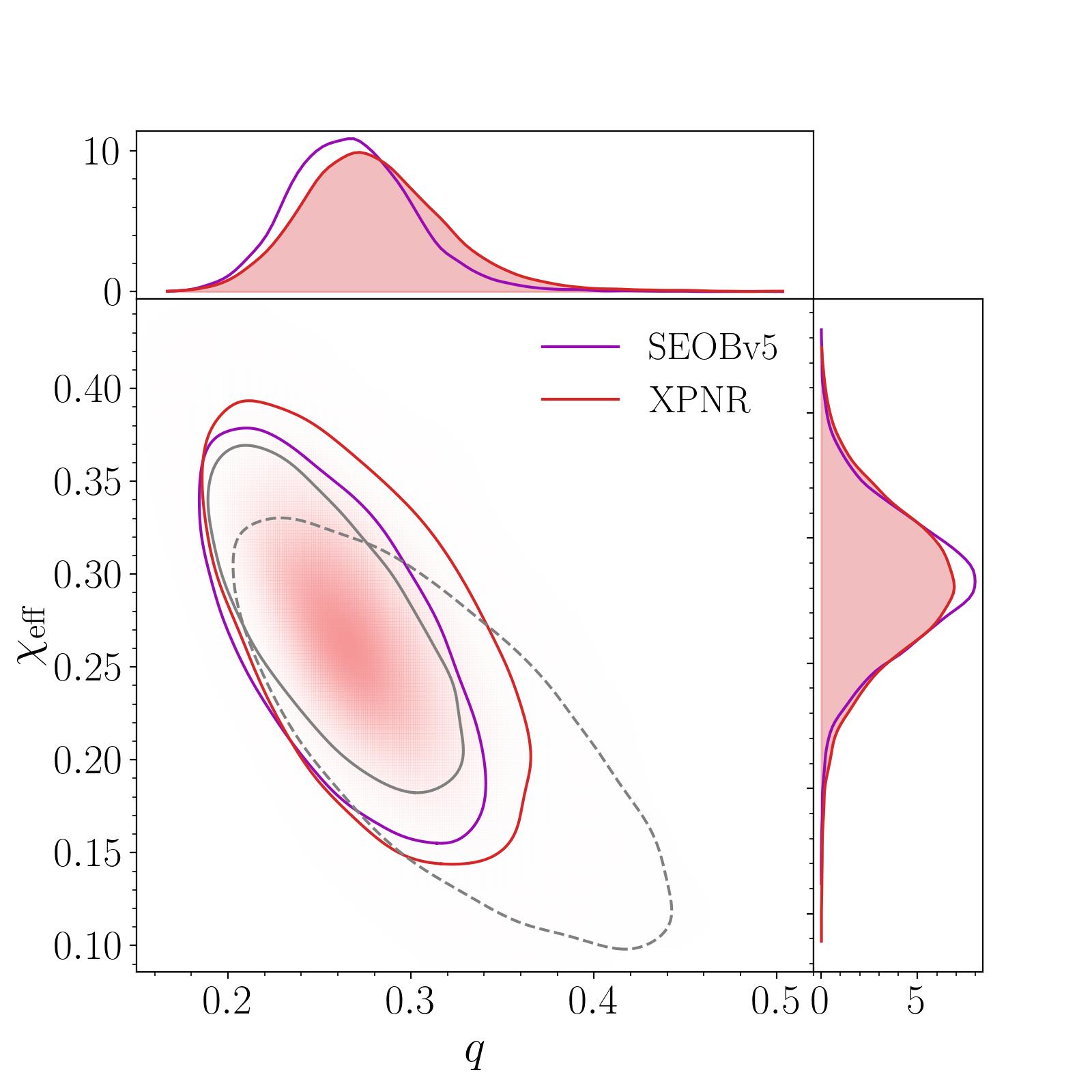} 
   \includegraphics[width=0.48\textwidth]{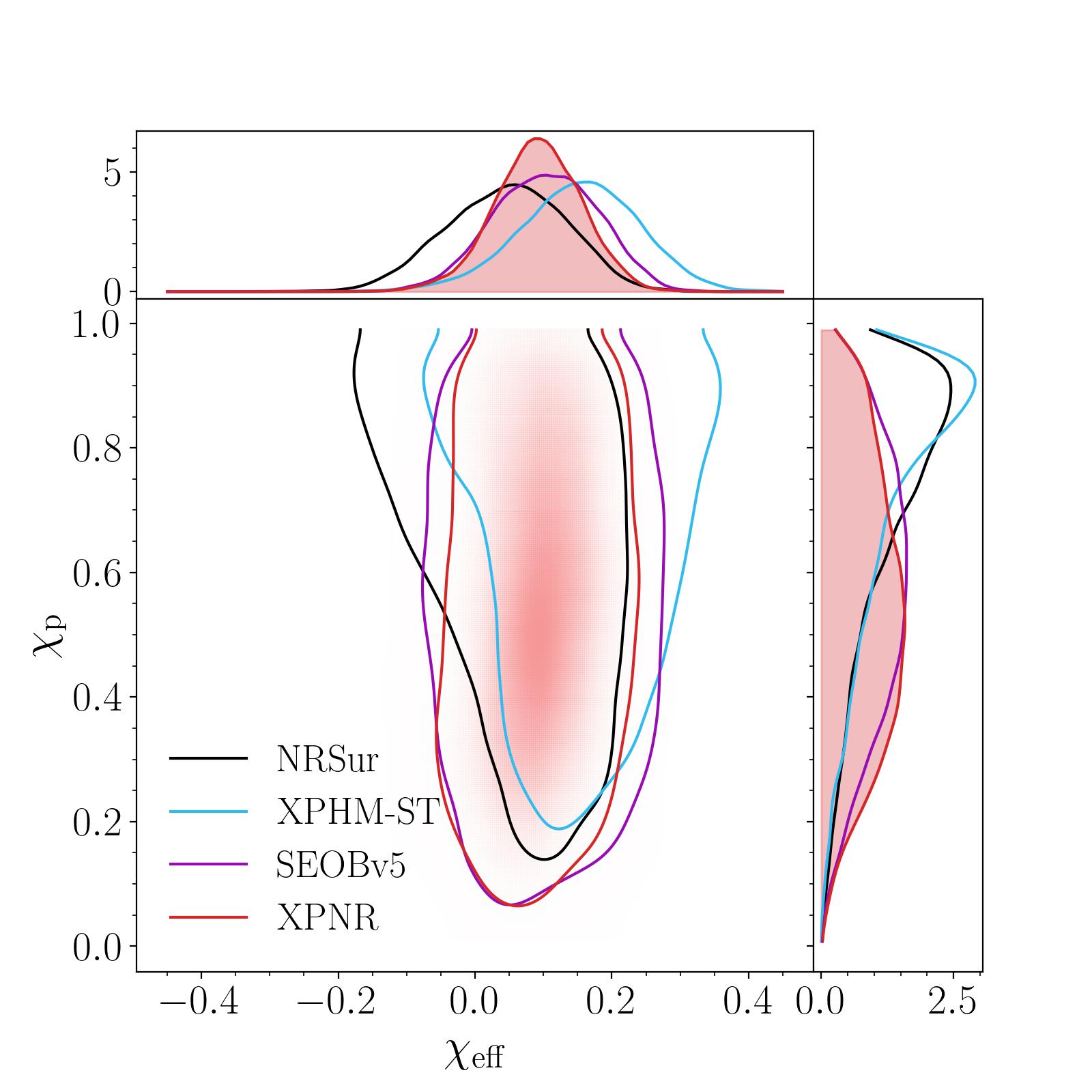}
   \caption{Comparison between posterior distributions obtained when analysing real gravitational wave events. In the \emph{Left} panel we show the two-dimensional marginalized posterior distribution for the mass ratio $q$ and effective parallel spin $\chi_{\mathrm{eff}}$ when analysing GW190412~\cite{LIGOScientific:2020stg}. We additionally show the \emph{Phenom PHM} and \emph{EOBNR PHM} posterior distributions reported in Ref.~\cite{LIGOScientific:2020stg} in grey dashed and grey solid respectively. In the \emph{Right} panel, we show the two-dimensional marginalized posterior distribution for the effective parallel spin $\chi_{\mathrm{eff}}$ and effective perpendicular spin $\chi_{\mathrm{p}}$ when analysing GW200129~\cite{KAGRA:2021vkt}. In both cases, the contours encase 90\% probability.}
   \label{fig:event_comparison}
\end{figure*}

Recently, Refs.~\cite{Kolitsidou:2024vub,Estelles:2025} showed that the evidence for precession in GW200129
increases when higher-order multipole asymmetries are included in the waveform model. Indeed, Ref.~\cite{Estelles:2025}
specifically showed that the inferred $\chi_{\mathrm{p}}$ more closely resembles the \textsc{NRSur7dq4}} result 
when higher-order multipole asymmetries are included in
{\textsc{SEOBNRv5PHM}}. Given that {\textsc{PhenomXPNR}} includes only the dominant
contribution to the multipole asymmetries, the smaller $\chi_{\mathrm{p}}$ measurement implies
that higher-order contributions may be needed to reproduce the high precession result.
This being said, {\textsc{PhenomXPHM-SpinTaylor}} does not include multipole asymmetries but still
measures a large $\chi_{\mathrm{p}}$. Interestingly, {\textsc{PhenomXPHM-MSA}} also
infers a large $\chi_{\mathrm{p}}$. We are unsure of the cause of this discrepancy.

Based on Bayesian analyses that we performed as part of this study, we see that
{\textsc{PhenomXPNR}} generally performs better than it's predecessor
{\textsc{PhenomXPHM-SpinTaylor}}. We understand that {\textsc{PhenomXPNR}} either
agrees well with {\textsc{SEOBNRv5PHM}} or performs better, particularly in the high mass
parameter space. This correlates with the mismatch results presented in Sec.~\ref{subsec:mismatches}.

\subsection{Timing tests}
\label{subsec:timing}

\begin{figure*}[htbp]
   \centering
   \includegraphics[width=0.48\textwidth]{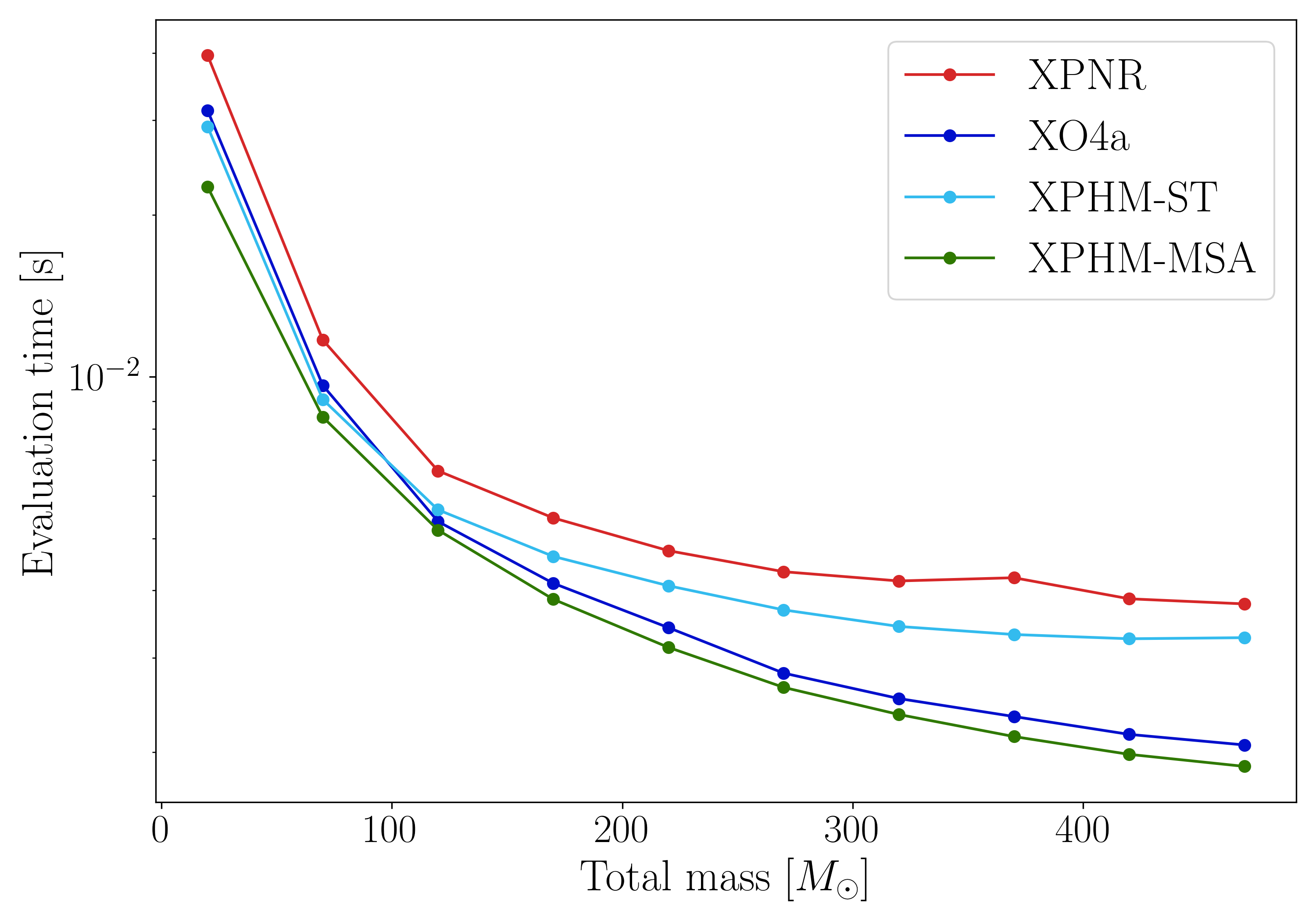}
   \includegraphics[width=0.48\textwidth]{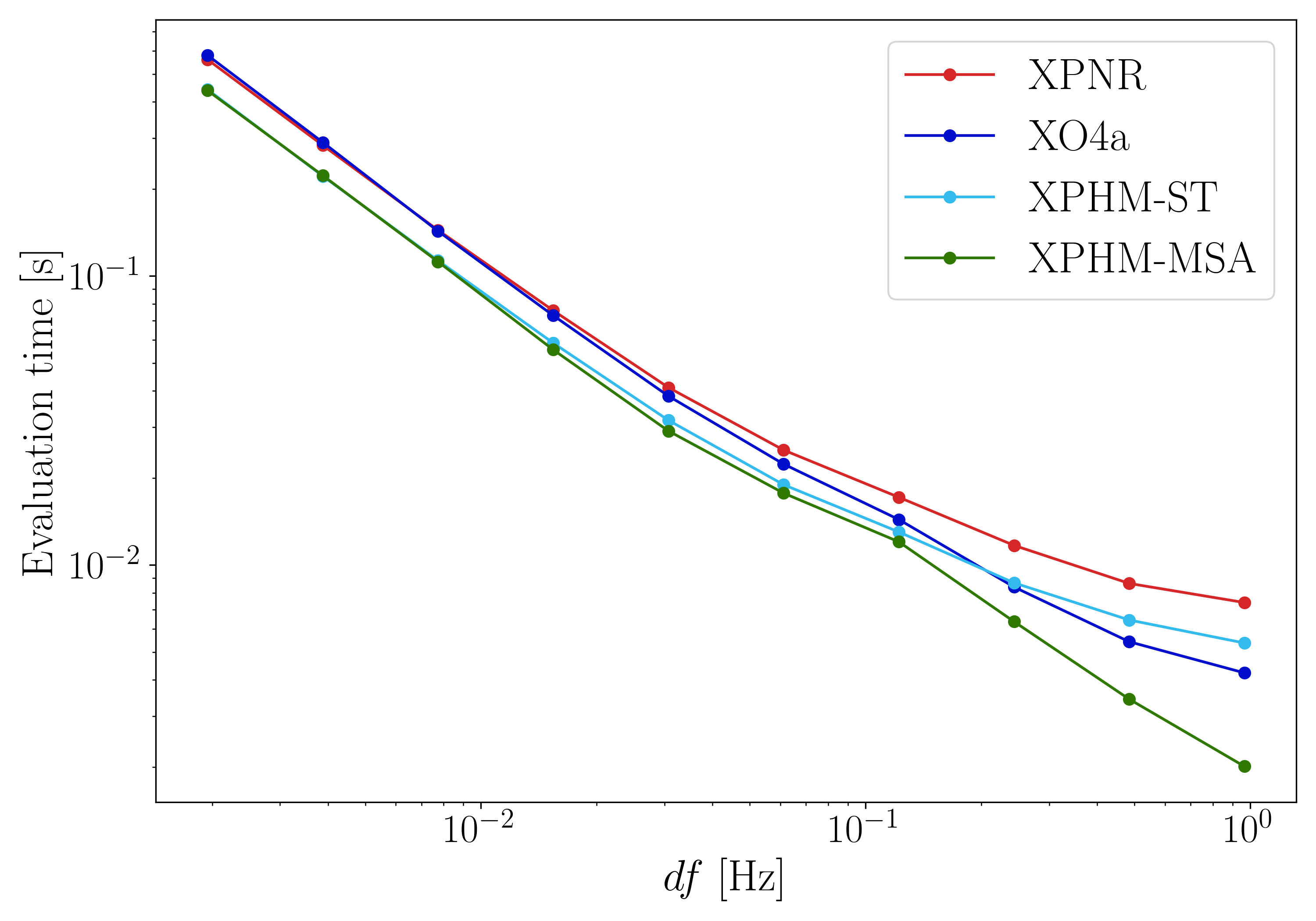}
   \caption{Waveform evaluation timing for four different \fd{}
   models: \textsc{PhenomXPNR}, \textsc{PhenomXO4}a, \textsc{PhenomXPHM-SpinTaylor} and \textsc{PhenomXPHM-MSA}.
   All timings were generated for a fixed binary configuration with 
   \(q=3\), \(\boldsymbol{\chi}_1=(0.2, 0, 0.3)\) and \(\boldsymbol{\chi}_2=(0, -0.5, -0.4)\)
   produced between $20$--$2048$~Hz. The left panel of the figure shows
   waveform evaluation times computed for a range of total masses between $20$--$500$~$M_\odot$
   with a fixed frequency interval \(df=0.125\)~Hz. The right panel displays runtime as a function
   of frequency spacing $df$ for a fixed total mass of $50$~$M_\odot$. All evaluation times are averaged
   over 50 waveform generations.}
   \label{fig:eval_time_scaling}
\end{figure*}

In this section we compare waveform evaluation times between multiple members of
the \textsc{Phenom} waveform family. We first compute the evaluation time 
for a fiducial binary configuration with \(Q=3\), \(\boldsymbol{\chi}_1=(0.2, 0, 0.3)\) and 
\(\boldsymbol{\chi}_2=(0, -0.5, -0.4)\), generated on an Apple M2 Max MacBook Pro laptop
across a frequency range
spanning $20$--$2048$~Hz and averaged over 50 waveform evaluations with identical parameters. This comparison
is performed both across a range of total masses, ranging between $20$--$500$~$M_\odot$
as shown in the left panel of Fig.~\ref{fig:eval_time_scaling}, and for a fixed
total mass of $50$~$M_\odot$ and across a range of frequency step sizes $df$, shown
in the right panel of the same figure. We compare the timing of four waveform models:
\textsc{PhenomXPNR}, \textsc{PhenomXO4}a, 
\textsc{PhenomXPNR-ST} and \textsc{PhenomXPHM-MSA}.

The results shown in Fig.~\ref{fig:eval_time_scaling} display a clear trend of increasing
waveform evaluation time across the four models that is largely independent of total
mass for this fiducial configuration, with \textsc{PhenomXPNR} being the slowest
model to evaluate and \textsc{PhenomXPHM-MSA} the fastest. We also observe that
at higher total masses, \textsc{PhenomXPHM-SpinTaylor} differs from both of the approximants
that use the inspiral MSA angles, and this timing performance is inherited by
\textsc{PhenomXPNR}. When \(df\) is varied, we see that for small frequency step sizes
the models separate into those that use the tailored frequency spacing 
for the inspiral precession angle interpolation,
detailed in Ref.~\cite{Thompson:2023ase}, and those which do not use this frequency
spacing for the precession angles.

We further investigate the timing performance of \textsc{PhenomXPNR} in comparison to
\textsc{PhenomXPHM-SpinTaylor} broadly across the two-spin precession parameter space, run
on the same hardware as the above analysis. We
randomly sample 10,000 points across \(Q\in[1,10]\), \(M\in[20,150]\)~\(M_\odot\),
and spin orientations drawn uniformly in \(\cos\theta_\text{LS}\in[-1,1]\), uniformly in
spin azimuthal angle between \([0,2\pi]\), and spin magnitudes ranging uniformly from~0 to~1. These
signals are generated across a frequency range of $20$--$1024$~Hz for sake
of memory and efficiency, and sampled at 
an appropriate frequency step size for the masses and spins
of the given random point, which we produce using an estimate of the signal 
chirp time from the \textsc{LALSimulation} function 
\texttt{XLALSimInspiralChirpTimeBound}. 

\begin{figure}[t]
   \centering
   \includegraphics[width=0.48\textwidth]{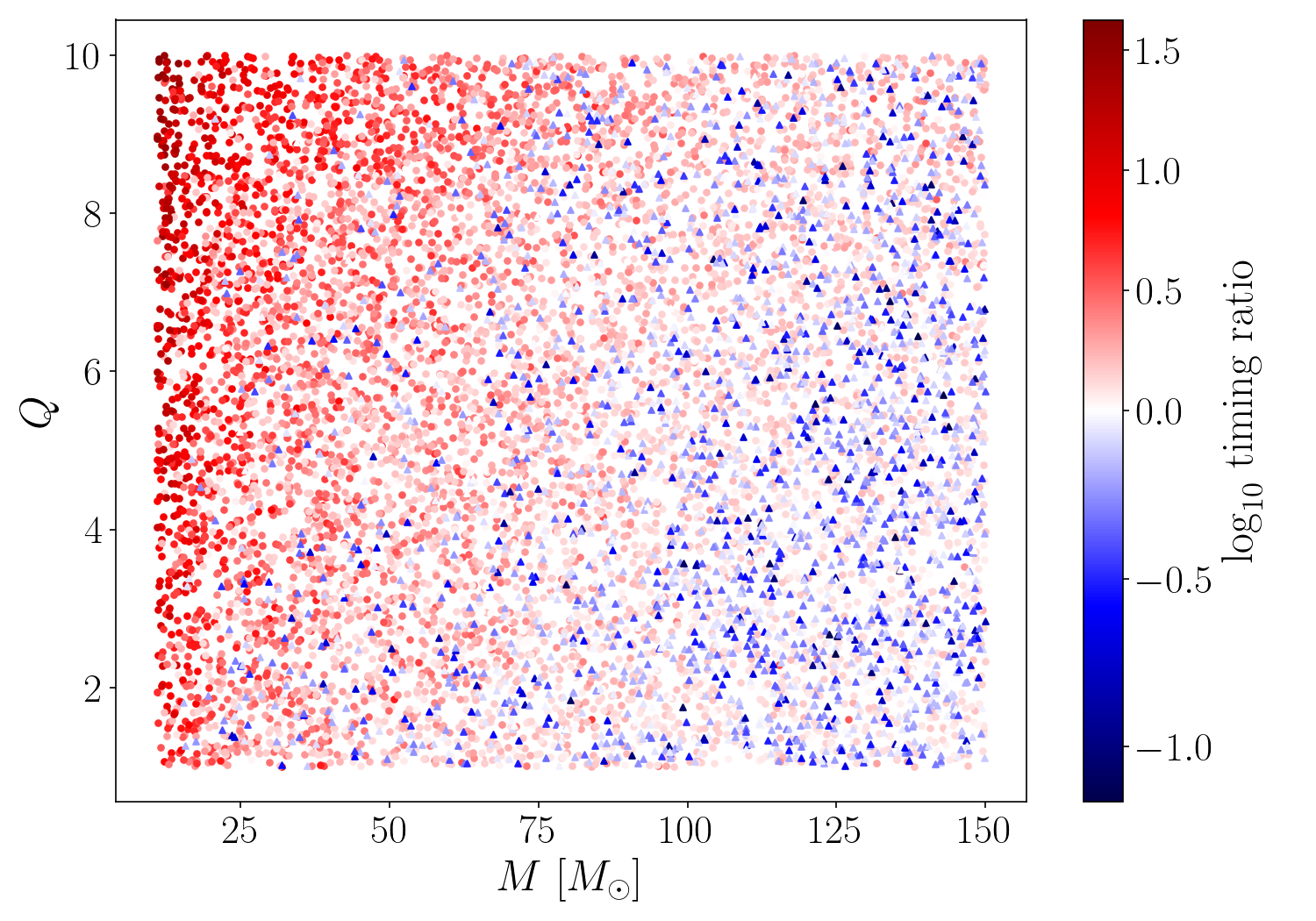}
   \caption{We compare the runtime evaluation between two models, 
   \textsc{PhenomXPNR} and \textsc{PhenomXPHM-SpinTaylor}, for a set of 10,000 points randomly
   drawn across the quasi-circular precessing parameter space. This figure shows
   the (\(\log_{10}\)) timing ratio of \textsc{PhenomXPNR} to \textsc{PhenomXPHM-SpinTaylor}
   plotted for the total mass \(M\) and mass-ratio \(Q\),
   with values greater than zero denoting a longer runtime for \textsc{PhenomXPNR} 
   compared to \textsc{PhenomXPHM-SpinTaylor} (the red circles) and values less than 
   zero denoting a faster runtime (the blue triangles).
   }
   \label{fig:runtime_ratio}
\end{figure}

\begin{figure}[t]
   \centering
   \includegraphics[width=0.48\textwidth]{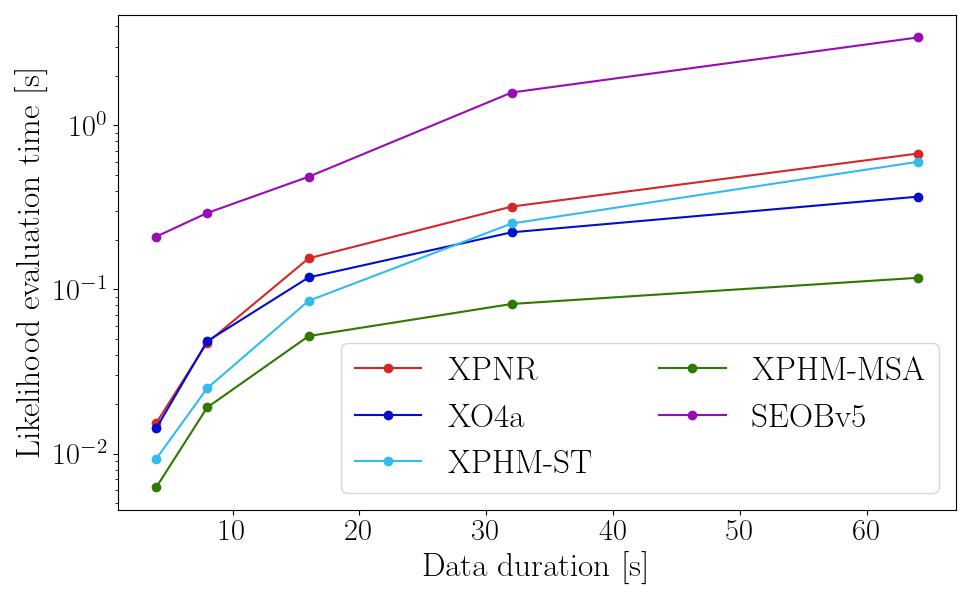} 
   \caption{Comparison between likelihood evaluation times for varying data duration. We compare the evaluation times for four different \fd{}
   models: \textsc{PhenomXPNR}, \textsc{PhenomXO4a}, \textsc{PhenomXPHM-SpinTaylor}, and the \td{} model: \textsc{SEOBNRv5PHM}. Each data point represents the mean of 10,000 likelihood evaluations calculated via {\textsc{bilby}}. Binary parameters were drawn from priors distributions that cut the chirp mass parameter space to ensure that the \gw{s} fit within the specified data duration: larger data durations correspond to smaller chirp mass binaries.}
   \label{fig:likelihood_timings}
\end{figure}

Figure~\ref{fig:runtime_ratio} shows the results for
the (\(\log_{10}\)) ratio of the timings for \textsc{PhenomXPNR} 
and \textsc{PhenomXPHM-SpinTaylor}, averaged over~3 waveform evaluations. The timing ratios 
show no dependence on any intrinsic
parameters except for \(Q\) and \(M\), so we restrict to plotting only
points across the total mass and mass-ratio range of evaluation. 
We identify timing ratios greater than one, \textit{i.e.,} 
when the evaluation of \textsc{PhenomXPNR} is slower than \textsc{PhenomXPHM-SpinTaylor},
with red circles in the plot, and blue triangles for points where the
timing ratio is less than one. From Fig.~\ref{fig:runtime_ratio} it is clear that
the runtime performance of \textsc{PhenomXPNR} depends most strongly on the total
mass of the system, with the transition mass across unity
in the timing ratio weakly depending on the mass-ratio. The slower waveform evaluation
at lower total masses arises from the finer frequency sampling used to generate
the interpolants for the precession inspiral angles to properly account for
two-spin oscillations, again 
detailed in Ref.~\cite{Thompson:2023ase}.

Finally we investigate the likelihood evaluation time of \textsc{PhenomXPNR} compared to 
multiple members of the \textsc{Phenom} waveform family. Given that the likelihood is
evaluated $O(10^{7})$ times in a typical Bayesian analysis, small differences in timing can
significantly impact the efficiency of a Bayesian inference analysis. We calculate the
evaluation time by taking the mean of 10,000 likelihood evaluations for different binary parameters
$\boldsymbol{\lambda}$ calculated via {\textsc{bilby}}; $\boldsymbol{\lambda}$ is randomly drawn from an agnostic prior distribution, with chirp mass cuts applied to ensure that the \gw{} signal fits within the specified data duration.
We enforce chirp masses between $25 \leq \mathcal{M} \leq 50$, $10 \leq \mathcal{M} \leq 25$, 
$6 \leq \mathcal{M} \leq 10$, $4 \leq \mathcal{M} \leq 6$, and $2 \leq \mathcal{M} \leq 4$ for the
$4$, $8$, $16$, $32$, and $64$ second data durations, respectively\footnote{Although the 32 and 64 second data durations contain binaries with masses consistent with the observed neutron star population, we ignored potential tidal affects and assumed that all binary components were black holes.}. Since we are uninterested
in the value of the likelihood, $d$ is randomly
generated Gaussian noise containing no signal. Each model evaluates the likelihood for
the exact same binary configurations. All likelihood evaluations were performed on
the same Lenovo AMD EPYC 7H12 node.

In Fig.~\ref{fig:likelihood_timings} we see that the likelihood evaluation time increases for
larger data durations. This is expected as the waveform evaluation time is longer for lower total mass binaries (see Fig.~\ref{fig:eval_time_scaling}), and larger data durations are needed to encase lower total mass binaries.
We see that for small data durations ($< 10$ seconds),
\textsc{PhenomXPNR} and \textsc{PhenomXO4}a have comparable likelihood evaluation
times, meaning that the extra physics added from the \textsc{PhenomXO4}a model dominates
the computational cost of \textsc{PhenomXPNR}. As the data duration increases, \textsc{PhenomXPNR}'s likelihood
evaluation time transitions from being dominated by the
 \textsc{PhenomXO4}a model, to being dominated by the \st evolution of the inspiral
 precession angles from \textsc{PhenomXPHM-SpinTaylor}. This transition occurs at around a data
 duration of 30 seconds. We note that for these longer data durations the whittle likelihood will 
 unlikely be used in Bayesian analyses; longer data duration analyses are more likely to use
 the Reduced Order Quadrature~\cite{Qi:2020lfr,Morisaki:2023kuq} or multibanded 
 likelihood~\cite{Morisaki:2021ngj} to reduce overall computational cost. The likelihood evaluation time for these
 long data durations are therefore unlikely to translate to the CPU time of typical analyses.
Across the whole space, \textsc{PhenomXPNR} is at most $1.9\times$
times slower than \textsc{PhenomXPHM-SpinTaylor} per likelihood evaluation. When averaging across all data durations considered,  \textsc{PhenomXPNR} is $1.5\times$ times slower than \textsc{PhenomXPHM-SpinTaylor}.
We also performed the same analysis with \textsc{SEOBNRv5PHM} and found that
\textsc{PhenomXPNR} is at least $3.1\times$ times faster than 
\textsc{SEOBNRv5PHM} per likelihood evaluation. Similarly, when averaging across all data durations considered, \textsc{PhenomXPNR} is $6.6\times$ times faster than 
\textsc{SEOBNRv5PHM}.

\section{Conclusions}
\label{sec:conclusions}

In this work, we have introduced \textsc{PhenomXPNR}, a new \gw{} model for \bbh{} mergers that accurately captures precessional effects in quasi-circular inspirals. Our new model seamlessly combines several efforts aimed at capturing precessional signatures in \gw{} signals: an accurate representation of the inspiral precession dynamics~\cite{Colleoni:2024knd}, a phenomenological description of the asymmetry between the $(2,2)$ and $(2,-2)$ multipoles of the \gw{} emission~\cite{Ghosh:2023mhc} and NR-calibrated models for the co-precessing frame multipoles and Euler angles accounting for precession effects in the merger-ringdown~\cite{Hamilton:2021pkf,Hamilton:2023znn}. 

We have provided a comprehensive overview of the different ingredients of the model in Sec.~\ref{sec:model}, where we have summarised the salient features of \textsc{PhenomXO4a}~\cite{Thompson:2023ase} and \textsc{PhenomXPHM-SpinTaylor}~\cite{Colleoni:2024knd}, as well as the technical aspects involved in blending them together.  

We then proceeded to provide quantitative measures of the enhanced performance of the model (Sec.~\ref{sec:performance}). In Sec.~\ref{subsec:angle_accuracy}, we demonstrated that the Euler angles predicted by \textsc{PhenomXPNR} better reproduce the ones extracted from NR simulations, significantly improving upon previous approximations. 
In Sec.~\ref{}, we have demonstrated that the inclusion of asymmetry in the dominant multipole consistently improves the accuracy of the model.
In Sec.~\ref{subsec:mismatches}, we have presented an extensive mismatch study against \textsc{NRSur7dq4}, showing that \textsc{PhenomXPNR} is currently the most accurate semi-analytical waveform model for face-on/off precessing quasi-circular binaries. We have then reinforced these conclusions through selected parameter estimation studies, outlined in Sec.~\ref{subsec:pe}. We analysed both simulated signals with known source properties and real gravitational wave events.
From the simulated signals, we have seen that, while all models can show biasses in certain regions of the parameter space, there are regions where, for high mass signals, the additional \nr{} calibration in \textsc{PhenomXPNR} enables it to correctly infer the correct source properties while models lacking this calibration exclude the correct values at 90\% confidence.
From the detected signals we can infer that \textsc{PhenomXPNR} performs reasonably when analysing real data.
Including additional physics or improving the accuracy of the current models increases the agreement between models for a given event, but current model differences still lead to discrepancies when analysing more challenging signals.
Finally, we have tested the performance of our model and compared it to previous precessing approximants, concluding that the improvements to model accuracy presented in this paper have resulted in a moderate decrease in efficiency.
However, we can see that \textsc{PhenomXPNR} is still appreciably more efficient than current state-of-the-art time domain models.

Overall, this model marks a significant step toward accurate and computationally efficient waveform models for precessing \bbh{} systems, providing a new tool to perform inference of spin-precession effects in \gw{} detections and probe the astrophysical origin and evolution of compact binaries. There are several key improvements that would further enhance the performance of \textsc{PhenomXPNR}, such as modelling the asymmetry between $\pm m$ modes for the full mode content of PhenomX, tuning the merger-ringdown Euler angles to two-spin NR simulations, including more accurate estimates for the final mass and spin of the remnant~\cite{Planas:2024vnq} and increasing the faithfulness of the model for sources with non-negligible inclination, by reassessing the modelling assumptions made for the higher harmonics of precessing systems in the \fd{}. Another potential extension of PhenomX templates would be the addition of other subdominant harmonics in the co-precessing frame, in particular of those carrying information about memory effects like the $(2,0)$ mode~\cite{Elhashash:2025hqi, Rossello-Sastre:2024zlr}. 

In the future, the insights gained in the construction of \textsc{PhenomXPNR} will be utilised to improve the precessing \td{} model \textsc{PhenomTPHM}. \td{} models offer a more intuitive picture of complex phenomena like precession or eccentricity and can greatly facilitate the construction of fully generic \imr{} templates including both~\cite{Albanesi:2025txj}; they can also find important applications in the context of data analysis for space-borne interferometry, where the use of \td{} response functions can naturally accommodate realistic LISA orbital information~\cite{Katz:2022yqe,Valencia:2025pgp}. 
However, the improved \fd{} model presented in this paper provides a highly accurate, computationally efficient model which can be employed robustly across the parameter space in order to face the challenges of gravitational wave astronomy with current generation ground-based detectors.

\section{Acknowledgements}

The authors would like to thank Hector Estelles, Jannik Mielke, Lorenzo Pompili, Shaun Swain and Thibeau Wouters for their efforts during the \texttt{IMRPhenomXPNR} code review. Additionally, we would like to thank Edward Fauchon-Jones for his contribution to the production of the BAM \nr{} simulations displayed in Fig.~\ref{fig: two spin beta}.

EH, MC, AH, JV, FARV and SH were supported by the Universitat de les Illes Balears (UIB); the Spanish Agencia Estatal de Investigaci\'on grants PID2022-138626NB-I00, RED2022-134204-E, RED2022-134411-T, funded by MCIN/AEI/10.13039/501100011033 and the ERDF/EU; Comunitat Auton\'oma de les Illes Balears through the Conselleria d'Educació i Universitats with funds from the European Union - NextGenerationEU/PRTR-C17.I1 (SINCO2022/6719) and from the European Union - European Regional Development Fund (ERDF) (SINCO2022/18146).

JT acknowledges support from the NASA LISA Preparatory Science grant 20-LPS20-0005. 
CH thanks the UKRI Future Leaders Fellowship for support through the grant MR/T01881X/1.
AH is further supported by grant PD-034-2023 co-financed by the Govern Balear and the European Social Fund Plus (ESF+) 2021-2027.
JV is additionally supported by the Spanish Ministry of Universities Grant No. FPU22/02211. 
FARV is also supported through the Conselleria d'Educació i Universitats via an FPI-CAIB doctoral grant (FPI\_092\_2022).
CGQ is supported by the Swiss National Science Foundation (SNSF) Ambizione grant PZ00P2\_223711.
SG was supported by the Max Planck Society’s Independent Research Group program.
LL acknowledges support at King's College London from the Royal Society {URF{\textbackslash}R1{\textbackslash}211451}. 
MH was supported by Science and Technology Facilities Council (STFC) grant ST/V00154X/1.

The catalogue of numerical simulations against which this model was calibrated, in addition go the simulations used for comparison in Fig.~\ref{fig: two spin beta}, were performed on the DiRAC@Durham facility, managed by the Institute for Computational Cosmology on behalf of the STFC DiRAC HPC Facility (www.dirac.ac.uk). The equipment was funded by BEIS capital funding via STFC capital grants ST/P002293/1 and ST/R002371/1, Durham University and STFC operations grant ST/R000832/1. In addition, several of the simulations used in the calibration were performed as part of an allocation graciously provided by Oracle to explore the use of our code on the Oracle Cloud Infrastructure. 

The authors are additionally grateful for computational resources provided by the LIGO laboratory and supported by National Science Foundation Grants PHY-0757058 and PHY-0823459, which were used to perform the match comparisons presented in this paper.

Parameter estimation and mismatch analyses were performed on a range of computing clusters. These include the Sciama High Performance Compute (HPC) cluster, which is supported by the ICG, SEPNet and the University of Portsmouth; the Picasso Supercomputer, with technical expertise and assistance provided by the SCBI (Supercomputing and Bioinformatics) center of the University of Malaga supported by grants AECT-2024-2-0017 and AECT-2025-1-0024; MareNostrum 5 at the Barcelona Supercomputing Center supported by grant AECT-2024-3-0027; the DiRAC Data Intensive service (DIaL) at the University of Leicester, managed by the University of Leicester Research Computing Service on behalf of the STFC DiRAC HPC Facility (www.dirac.ac.uk). The DiRAC service at Leicester was funded by BEIS, UKRI and STFC capital funding and STFC operations grants. DiRAC is part of the UKRI Digital Research Infrastructure.

Further analyses were performed on the supercomputing facilities at Cardiff University operated by Advanced Research Computing at Cardiff (ARCCA) on behalf of the Cardiff Supercomputing Facility and the HPC Wales and Supercomputing Wales (SCW) projects. We acknowledge the support of the latter, which is part-funded by the European Regional Development Fund (ERDF) via the Welsh Government. In part the computational resources at Cardiff University were also supported by STFC grant ST/I006285/1.

This research has made use of data or software obtained from the Gravitational Wave Open Science Center (gwosc.org), a service of the LIGO Scientific Collaboration, the Virgo Collaboration, and KAGRA. This material is based upon work supported by NSF’s LIGO Laboratory which is a major facility fully funded by the National Science Foundation, as well as the Science and Technology Facilities Council (STFC) of the United Kingdom, the Max-Planck-Society (MPS), and the State of Niedersachsen/Germany for support of the construction of Advanced LIGO and construction and operation of the GEO600 detector. Additional support for Advanced LIGO was provided by the Australian Research Council. Virgo is funded, through the European Gravitational Observatory (EGO), by the French Centre National de Recherche Scientifique (CNRS), the Italian Istituto Nazionale di Fisica Nucleare (INFN) and the Dutch Nikhef, with contributions by institutions from Belgium, Germany, Greece, Hungary, Ireland, Japan, Monaco, Poland, Portugal, Spain. KAGRA is supported by Ministry of Education, Culture, Sports, Science and Technology (MEXT), Japan Society for the Promotion of Science (JSPS) in Japan; National Research Foundation (NRF) and Ministry of Science and ICT (MSIT) in Korea; Academia Sinica (AS) and National Science and Technology Council (NSTC) in Taiwan.

\bibliography{paper.bib}

\end{document}